\documentclass[10pt]{article}
\usepackage{amsfonts,bbm,amsmath}
\usepackage{epsfig,graphics,mathabx,graphicx,mathrsfs,amsthm}
\usepackage{eepic,bm,epsfig,color,subfigure}
\DeclareMathAlphabet{\mathpzc}{OT1}{pzc}{m}{it}
\def\bra#1{\mathinner{\langle{#1}|}}
\def\ket#1{\mathinner{|{#1}\rangle}}

\def\ave#1{\langle #1\rangle}

\def\comp{\mathbb{C}}
\def\genX{\hat{x}}
\def\genZ{\hat{z}}

\def\vect#1{\tt #1}
\def\heis#1{{\bf H}_{#1}}
\newcommand{\pj}[1]{\ket{#1}\bra{#1}}
\newcommand{\prj}[2]{\ket{#1}\bra{#2}}
\newcommand{\inp}[2]{\mathinner{\langle #1|#2\rangle}} 
\newcommand{\inpr}[3]{\mathinner{\langle #1|#2|#3\rangle}} 
\newcommand{\inpra}[2]{\mathinner{\langle #1|#2|#1\rangle}} 
\newcommand{\inprab}[2]{\mathinner{\bigl\langle #1\bigl|#2\bigr|#1\bigl\rangle}} 
\newcommand{\beq}{\begin{equation}}
\newcommand{\eeq}{\end{equation}}
\newcommand {\cali}[1]{{\mathcal #1}}
 
\newcommand{\diffD}{{\mathrm d}}

\newcommand{\tr}{{\tt Tr}} 
\newcommand{\hconj}[1]{{#1}^{\dagger}} 
\newcommand{\conj}[1]{\overline{#1}}
 
\newcommand{\pmat}{\begin{pmatrix}} 
\newcommand{\emat}{\end{pmatrix}} 
\newcommand{\complex}{\mathbb{C}}
\newcommand{\real}{\mathbb{R}}

\newcommand{\norm}[1]{|\!|#1|\!|} 
\newcommand{\tensor}{\otimes}

\newtheorem{propn}{Proposition}{}{}
\newtheorem{thm}{Theorem}{}{}
\newtheorem{lem}{Lemma}{}{}
{}{}
{}{}
{}{}
{}
\newcommand{\commentout}[1]{}
\newcommand{\be}{\begin{enumerate}}
\newcommand{\ee}{\end{enumerate}}
\newcommand{\bi}{\begin{itemize}}
\newcommand{\ei}{\end{itemize}}
\newcommand{\floor}[1]{\lfloor #1\rfloor}
\newcommand{\bfloor}[1]{\bigl\lfloor #1\bigr\rfloor}
\newcommand{\intg}{{\mathbb Z}}
\newcommand{\sgn}{\operatorname{sgn}}
\begin{document}
\title{Quantum Fourier transform, Heisenberg groups and quasiprobability distributions} 
\author{Manas K. Patra and Samuel L. Braunstein   \footnote{{\bf email}:\{manas,schmuel\}@cs.york.ac.uk}\\
Department of Computer Science \\ University of York, York YO10 5DD, UK}
\date{}
\maketitle
\begin{abstract}
This paper aims to explore the inherent connection among Heisenberg groups, quantum Fourier transform and (quasiprobability) distribution functions. Distribution functions for continuous and finite quantum systems are examined from three perspectives and all of them lead to Weyl-Gabor-Heisenberg groups. The quantum Fourier transform  appears as the intertwining operator of two equivalent representation arising out of an automorphism of the group. Distribution functions correspond to certain distinguished sets in the group algebra. The marginal properties of a particular class of distribution functions (Wigner distributions) arise from a class of automorphisms of the group algebra of the Heisenberg group. We then study the reconstruction of Wigner function from the marginal distributions via inverse Radon transform giving explicit formulas. We consider applications of our approach to quantum information processing and quantum process tomography. 
\end{abstract}
\section{Introduction}
Quasiprobability distribution functions (or simply distribution functions) on a quantum system provide an alternative and equivalent description of quantum states. We will discuss three possible approaches to distribution functions. The first approach is essentially Wigner's original approach \cite{Wigner32} and it attempts to give a ``phase-space'' description of quantum states. 
The state of a quantum system determines the probability distributions of its observables. It is possible to completely specify the state by giving the distributions of functions of a pair of {\em conjugate} nondegenerate observables $\hat{f}$ and $\hat{g}$. The most well-known example is the position-momentum pair. Thus corresponding to a (mixed) state $\rho$ we associate a {\em real} function $W(p,q:\rho)$ of ``$c$-number'' variables which have the same information content as the state and give the correct marginals. Now for conjugate observables the expectation values of the functions $\phi(\hat{f},\hat{g})$ will specify the state completely (ignoring the questions of operator ordering). In classical probability theory these expectation values are generated by the {\em characteristic function}. The characteristic function of (classical) random variables $X_1,X_2,\dotsc,X_n$ with joint  probability distribution $F(x_1,\dotsc,x_n)\equiv F({\bf x})$ is given by \cite{Shiryayev}
\[
 \widetilde{F}({\bf t}) =  \int e^{i{\bf t}\cdot{\bf x}} \mathrm{d}F({\bf x})
\]
where ${\bf t}\cdot{\bf x}$ is the usual Euclidean scalar product of two real vector ${\bf x}= (x_1,\cdots, x_n)^T$ and ${\bf t}= (t_1,\cdots, t_n)^T$ where $T$ denotes transpose. If the probability distribution is given by a probability density $p({\tt x})$ then the characteristic function is simply the Fourier transform of $p({\tt x})$. Another way of viewing the characteristic function is to note that $\tilde{F}({\bf t})=\ave{e^{i{\bf t}\cdot{\bf X}}}$, the expectation value of the complex random variable $ e^{i{\bf t}\cdot{\bf X}}$. Then assuming the existence of a probability density it is given by the inverse Fourier transformation 
\beq \label{eq:inv1}
p({\bf x})=\int e^{-i{\bf t}\cdot{\bf x}}\ave{e^{i{\bf t}\cdot{\bf X}}}{\mathrm d}{\bf t}
\eeq
In the form \eqref{eq:inv1} it is suitable for a ``quantum'' extension \cite{Moyal}. Of course, in the quantum case the observables (or quantum random variables) $X_i$ will {\em not} commute in general and we have the problem of interpreting the function $p$ as a joint probability distribution. However for a set of {\em compatible} or commuting observables a joint distribution is unambiguously defined. For incompatible observables we may take \eqref{eq:inv1} as the definition of joint probability distribution. 
The values that are obtained on joint measurement of these observables constitute the joint spectrum which, in general, may have both continuous and discrete segments. In the finite-dimensional case we have a finite spectrum and hence the integral has to be replaced by a sum. But there are problems in interpreting this as a function in a classical phase space \cite{Strat}. Alternatively, we can work with {\em finite} Fourier transform. Such a transform is defined over a finite abelian group. From the structure of such groups we may focus on the group 
$Z_N=\intg/N\intg$, the additive group of integers modulo $N$. We can thus take our ``configuration space'' to be $Z_N$. This in turn forces us to consider observables which take ``values'' in the set $\{ 0, 1, \dotsc, N-1\}$ where $N$ must now be identified with the dimension of the Hilbert space. Since we are concerned only with probability distributions of the possible outcomes this is really not a restriction. Operationally, we can always calibrate our instruments to yield these outcomes. The ``position'' and  ``momentum'' variables are both identified with $Z_N$ and the corresponding unitary representations respectively act multiplicatively and additively on the ``position space''. So we have two representation of $Z_N$. But they do not constitute a representation of the ``phase space'' $Z_N\times Z_N$ as the latter is a commutative group. The simplest possible {\em noncommutative} extension is a central extension \cite{Rotman}. After some restrictions due to finiteness of the dimension we get a Heisenberg group. 

Now let us approach the problem of distribution function from another perspective. The state of the system, $\rho$ is a positive definite operator with trace 1. In infinite dimensions they belong to a special class called {\em trace class} operators. In finite dimensions every operator is clearly trace class. Trace class operators admit a Hilbert space structure. Thus for two such operators $A,B$ define $(A,B)=\tr(\hconj{A}B)$. In finite dimensions this introduces the familiar Frobenius or Hilbert-Schmidt norm. If we restrict to hermitian operators we get a real Hilbert space $\cali{K}$. Pick an orthonormal basis $\{A_i: i=1,2,\dotsc, \}$ in $\cali{K}$. We can write the state 
\[\rho = \sum_i W_i(\rho)A_i\text{ with } W_i(\rho)= \tr(\rho A_i) \]
Clearly $W_i$ are real. If we also demand that $\sum_iW_i=1$ then we have a quasiprobability distribution over the index set $\mathcal{I}\;(i\in \mathcal{I})$. We want this index set to have a classical interpretation and a natural choice is the {\em phase space}. Then, $i=(x,z)$ is a {\em pair} of indices\footnote{We often use $z$ in place of $p$ for momentum variable to avoid confusion with probability density. Further, $x$ and $z$ can be vectors.}. Henceforth, we assume this and write $A(x,z)$ instead of $i$. The works \cite{Baker, Wootters87} follow this approach to distribution function (see also \cite{Vourdas} for a review in the finite-dimensional case). 
Thus the choice of distribution is equivalent to choice of special bases. The operators $A(x,z)$ are called phase-point operators. Actually, they have to satisfy some extra conditions. We will see that the phase point operators correspond to certain sets (called Wigner sets) in the group algebra of the Heisenberg group. 

There is yet another view of distribution function which has origins in signal analysis. A signal may be represented in the time domain as $f(t)$ or {\em frequency} domain as $\tilde{f}(\omega)$ (Fourier transform). Thus we represent the signal in terms of ``elementary'' harmonic signals and the coefficients give the representation in frequency domain. But it can also be represented by other elementary non-harmonic signals with minimum uncertainty. This was Gabor's seminal idea \cite{Gabor46}. Unlike harmonic signals Gabor's elementary signals are localized in time {\em and} frequency domains. This joint time-frequency domain is the analogue of phase space. How do we generate these elementary signals? Starting with a ``reasonable'' initial signal say a Gaussian function in the time domain we apply a sequence of two operators, multiplication {\em and} translation, which are ``diagonal'' in the time and frequency domain respectively. The resulting sequence of functions are used to represent an arbitrary signal. Now to represent a vector in some space we don't need a basis, any set of vectors that {\em span} the space will do. In finite dimensional Hilbert space such sets define {\em frames} \cite{Christensen03}. An example of such overcomplete sets is the set of coherent states in quantum optics. The frame-theoretic approach to distribution functions was recently proposed in \cite{Ferrie}.  We will not go into details of frame theory but mention that frames are a generalization of orthonormal bases in Hilbert space. In this context, Gabor's elementary functions constitute Gabor-Weyl-Heisenberg (GWH) frames. Most distribution functions are examples of such frames although there are some exceptions. The GWH-frames are generated by applying sequence of translation and multiplication operators to (continuous) signals creating a function in the time-frequency domain (the phase space!). These operators generate a discrete Heisenberg group. 

Finally, we come to another significant property of the Wigner function, a particularly important distribution function. Let $W(x,p)$ be the continuous Wigner function with $x$ and $p$ representing the classical position and momentum variables. Then the marginals $\sum_x W(x,p)$ and $\sum_p W(x,p)$ are {\em probability} distributions of the {\em quantum} momentum and position operator respectively. Further, if we sum $W(x,p)$ along some line $ax+bp=0$ then the resulting function is the probability distribution of a quantum operator ``orthogonal'' to $c\hat{x}+ d\hat{p}$ where $(c,d)$ is a vector orthogonal to $(a,b)$ in the $x-p$ plane. We will make these definitions precise later. Call this the Radon property. This is an important property and  can be used to invert the transform. We will see that the Radon property is related to the transformation properties of distribution function under some automorphisms of the Heisenberg group. We note that the Radon property is very useful in the practical problem of reconstruction of states and processes. We prove a general Radon property which gives us a lot of freedom in our choice of possible measurements. 

The brief (and incomplete) survey of the approaches to distribution functions in the preceding paragraphs indicates that a lot of work has been done in this area\footnote{See \cite{Ferrie},  \cite{Durt} and \cite{Vourdas} for large lists of references relevant to the present work.}. Besides their theoretical significance distribution functions have application in state tomography \cite{Leonhardt95}, statistical mechanics and quantum optics \cite{Leonhardt97}. It is also intimately connected with the theory of coherent states. The GBH type  operators after complexification and some algebra give rise to the familiar {\em displacement} operators. The coherent states are the orbits of Weyl-Heisenberg group (henceforth only Heisenberg group)\cite{Perelomov}. In this work we mainly focus on the finite-dimensional case. This case presents some difficulties absent in the continuous case. The finite dimensional case is also  significant for quantum information processing \cite{Miquel,Paz05}. 

We have considered $Z_N\times Z_N$ as the basic model of finite ``phase space''. In the literature other phase spaces have been considered (see \cite{Ferrie} for a discussion and references). The intrinsic structure of these phase spaces may have interesting bearing on the corresponding distribution functions. In particular, some authors have considered finite field $F_N$ with $N$ elements instead of $Z_N$ (see e.g.\ \cite{Gibbons}). This is only possible if $N=p^n$ is a power of some prime $p$. Now $Z_p$ and $F_p$ coincide. In general, the additive groups of $F_{p^n}$ and $Z_p\times \dotsb \times Z_p$ ($n$ factors) are isomorphic. We can consider the Heisenberg groups over the latter (by central extension). We do not follow this here as the paper is already quite large. However, note that the  analogy between what the authors in \cite{Gibbons} call quantum nets and Wigner sets defined in this paper. More precisely, quantum nets correspond those Wigner sets which are permuted by the action of the automorphism group $SL(2,Z_N)$. 
The paper is quite self-contained. We give most of the proofs. Some of the results are known but were derived using different approaches. Let us first note some of the main contributions of the present work. 
\be
\item
We use the Heisenberg groups as the basic approach to distribution functions. As we have seen in the preceding paragraphs this is {\em the} unifying thread tying the different approaches and perspectives on distribution function. 
\item
Heisenberg group has been used in the literature context of distribution functions. But here we use discrete Heisenberg groups and family of finite quotient groups thereof. We define these groups abstractly in terms of generators and relations. Thus we can consider different representations (irreducible and reducible) and operators between representations. 
\item
Our treatment of the Heisenberg groups is defined in terms of generators and relations. This makes the computations and proofs easier. Further, we don't need the language of (pseudo) phase spaces. 
\item
We show that the existence of distribution functions is equivalent to certain sets in the group {\em algebras} of the Heisenberg groups. This provides us with powerful methods of representation theory. We list some of the outcomes by the use of these methods. 
\be
\item
The analysis of marginal distributions become transparent. They correspond to an invariance up to permutations under certain groups of automorphisms (e.g.\ $SL(2,Z_N)$) of the Heisenberg groups. 
\item
Demanding this invariance we get unique distribution functions in odd dimensions. 
\item
We also infer that it is impossible to retain full invariance {\em and} other properties like hermiticity and linear independence even dimension. We therefore have three possible strategies: i. drop the requirement of invariance, ii. drop the requirement of independence or hermiticity or iii. require invariance under a smaller set of transformations. We discuss all three and give some alternative candidates for distribution functions in even dimensions.  
\item
Our analysis via the automorphism groups obviates the need for ad hoc hypothesis and guess work. 
\ee
We note again that although the results mentioned in (a), (b) and parts of (c) are known our approach via automorphisms is different. 
\item
We give explicit formulas in most cases. The close connection with finite Radon transform is made clear. It is used to derive the formulas for state reconstruction. The inversion formulas in the case o dimension $=2^n$ the formulas appear to be new. 
\item
We explore applications to quantum computing and information. The fact that Weyl-Heisenberg groups describe the kinematics of quantum systems is known \cite{Weyl, Schwinger70}. We show that the {\em dynamics} is described by (unitary) automorphisms of the {\em group algebra}. The case of unitary automorphisms of the {\em group} itself is analyzed in \cite{Appleby}. The latter correspond to the Clifford group and to go beyond it (``non-classical'' dynamics) we have to consider the group algebra. We also illustrate the utility of the Heisenberg groups in quantum circuits. We give an application to quantum process tomography. 
\ee

We now give a summary of the paper. 
In Section \ref{sec:QFT} we discuss the quantum Fourier operators. The quantum Fourier transform (QFT) is another form of finite Fourier transform \cite{Terras}. Consider the two representations of $Z_N$ acting as multiplication and translation respectively. The QFT connects the two. The generators of the two representations give us the basic operators: $Z$ and $X$. We can consider these as the finite-dimensional analogue of unitary operator generated by ``position'' and ``momentum'' respectively. In Section \ref{sec:quasprob} we review (continuous) quasi-probability distribution functions or simply distribution functions. The continuous distribution functions are somewhat easier to deal with because the ``infinitesimal'' generators satisfy simple commutation rules (the Heisenberg relations).  In Section \ref{sec:di_quasiprob} we come to one of our main themes the finite distribution functions. We list a set of properties, satisfied by the continuous Wigner function, and demand that {\em any} distribution function must satisfy them. In particular, we give examples of discrete Wigner functions. Here we encounter the difficulties when the dimension is even. We also derive explicit formulas for the phase-point operators. The odd-dimensional case (apart from some constants) is essentially same as Wootters' \cite{Wootters87} operators in prime dimension. 

In Section \ref{sec:Heis} we study the Heisenberg groups. We start with the continuous version as it has been studied well in connection with Fourier transforms \cite{Folland89}. We then look at discrete and finite Heisenberg groups, their structure, representation and automorphisms, all of which play important role in our study of the finite distribution functions. We show that there is a one-to-one correspondence between distribution functions in dimension $N$ and certain sets $\{A(x,z)\}$, (called Wigner sets) in the group algebra of the Heisenberg group $\heis{}$ in that dimension. The representation of these sets are the phase-point operators. A slight generalization of the Wigner sets may be used to define Weyl-Heisenberg frames. The group $SL(2,Z_N)$ of $2\times 2$ matrices in $Z_N$ with determinant 1 induce automorphisms on the Heisenberg group $\heis{}$. Thus for each $M\in SL(2,Z_N)$ we define an automorphism $\sigma_M$ of $\heis{}$. These automorphisms, in turn, determine the marginal properties of the Wigner functions. Thus if $W(x,z)\equiv W(\bm{\zeta})$ is a distribution function then the functions  
 \[   Q(z) =\sum_x W(M^{-1}\bm{\zeta})\text{ and } P(x)= \sum_z W(M^{-1}\bm{\zeta})\]
 are the marginals. $Q(z)$ is called a simple marginals if it is the probability distribution (in the given state)  of an observable $\hat{z}_M$ defined by $e^{i\hat{z}_M}=\sigma_M(Z)$. A similar definition can be given for $P(x)$. We show the necessary and sufficient condition for the existence of simple marginals for all members of $SL(2,Z_N)$. Thus the requirement that all marginals be simple determine the distribution function up to isomorphism. In the case of odd dimensions for the Wigner function all marginals are simple. This can be neatly expressed as follows. Let $A_M(x,z)=A(M^{-1}x,M^{-1}z)$. Then $\{A_M(x,z)\}$ is also a Wigner set. An analogous result is called Clifford invariance in \cite{Gross}. 
This is not true in even dimensions. We investigate three alternatives by weakening our requirements. First, we do not demand that the phase-point operators form a basis. Now they constitute a {\em frame}. This is the most common approach (see for example,  \cite{Leonhardt96,Miquel}). The marginal conditions are simple but at the expense of losing orthogonality of bases.  We show that this is similar to the case of spin-1/2 representation in the sense that a complete ``rotation'' does not preserve the values of the functions involved. More precisely, we get functions which are not periodic on $Z_N$. However, they have period $2N$. So we go over to $Z_{2N}\times Z_{2N}$ as the phase space. 
Next we drop the requirement that the marginals be simple in the above sense. It is still possible to compute the marginals in terms of the probability distribution of the observable $\hat{z}_M$. We indicate how explicit formulas can derived to compute this. Finally, since it is not possible to satisfy simple marginal conditions on all of $SL(2,Z_N)$ we consider certain subsets adequate for inversion, that is, computing the state from the marginal data. We give such a subset in dimension $N=2^K$. A similar construction from a different perspective was done in \cite{Chaturvedi09} but we present our formulas in an explicit functional form. 

In the section \ref{sec:inverseRadon} we explore the fact that the definition of marginals is a Radon transform of $W$ in the sense of \cite{Diaconis,Fill,Velasquez}. We then give the inversion formulas in several cases. The inversion formulas for odd dimensions were given in \cite{Leonhardt95}. Our derivation, however, is more general and applicable to {\em any} finite distribution function. The important point is we can invert these transform and recover the Wigner function and hence the quantum state. In the next section we discuss some applications to quantum information processing. We provide some simple relations between standard quantum gates and operators representing the Heisenberg groups which will prove useful for implementing the state and process determination schemes using phase-point operators. We give formulas for quantum process tomography using phase-point operators. In the final section we discuss some more possible applications and future work. 

\section{Quantum Fourier Operators}\label{sec:QFT}
Let $G={\mathbb Z}/N{\mathbb Z}$ be the additive group of integers modulo-$N$. There are two obvious representations of $G$ on an $N$-dimensional Hilbert space $H$. Let $g$ be a generator of $G$.  Suppose $\phi:G\rightarrow \cali{U}(H)$ is faithful representation of $G$ by unitary operators where $\cali{U}(H)$ is the set of unitary operators on $H$. If $\phi(g)=Z$ then we must have $Z^N=1$ since the order of $G$ is $N$ and the representation is faithful. The eigenvalues of $Z$ are $N$th roots of unity. Let $\{\ket{i}:i=0,\cdots,N-1\}$  be the corresponding eigenvectors such that $Z\ket{i}=\omega^i\ket{i}$ where $\omega$ is a primitive $N$th root. Call it the the {\em computational basis} $\cali{B}_c$. There is another representation $\phi'$ of $G$ defined by $\phi'(g)=X$ where $X\ket{i}= \ket{i+1\pmod N}$. We can think of $\phi$ as the multiplicative and $\phi'$ as the additive representations. $Z$ and $X$ represent multiplication and translation operators resp. Clearly, $X$ is unitary and there is basis $\cali{B}_f$ in which it is diagonal. The quantum Fourier transform (QFT) is the unitary map connecting the two representations taking $\cali{B}_c\rightarrow \cali{B}_f$. The eigenvalues of $X$ are also roots of unity as $X^N=I$. Since $\phi'$ is also faithful the diagonalization of $X$ yields $Z$ fixing the ordering. Hence there exists a unitary operator $\Omega$ such that
\beq \label{eq:basicReln}
\hconj{\Omega}X\Omega = Z
\eeq
The explicit form of $\Omega$ in the computational basis is easy to compute. Thus, if $\alpha=\sum_i x_i\ket{i}$ is and eigenvector then $X\alpha= u\alpha$ implies \(
 x_0=u x_1,\dotsc, x_{N-2}= u x_{N-1},\) and $ x_{N-1}= u x_0$. 
This yields after normalization
\beq\label{eq:fourierMat}
(\Omega)_{ij} = \frac{1}{\sqrt{N}} \omega^{-ij}
\eeq
So the quantum Fourier transform (QFT) is the map 
\[ \ket{k} \rightarrow \frac{1}{\sqrt{N}} \sum_j \omega^{-jk}\ket{j} =\frac{1}{\sqrt{N}} \sum_j e^{-2\pi ijk/N}\ket{j} \]
We note that we follow the convention of mathematicians in the definition of discrete or finite Fourier transform. In the quantum information literature the usual definition is with a positive sign in the exponent which is our {\em inverse} Fourier transform. Now $X$ and $Z$ can be expressed as 
\beq \label{eq:gen}
X=e^{i\hat{x}} \text{ and } Z=e^{i\hat{z}}
\eeq
where $\hat{x}$ and $\hat{z}$ are the hermitian generators of the respective unitary rotation. Moreover, their eigenstates are (discrete) Fourier transforms of each other. This is reminiscent of position and momentum observables which also have the property that their (generalized) eigenstates are (continuous) Fourier transforms of each other. We may therefore regard the observables $\hat{x}$ and $\hat{z}$ as conjugate ``dynamical variables''. This terminology is further justified by the following observation which is crucial for our calculations of quasi-probability distributions. 
\beq\label{eq:commut} 
XZ = \omega ZX
\eeq
This is most easily derived by applying both sides to vectors in the computational basis $\cali{B}_c$. We observe that the unitary operators $e^{ia\hat{p}} \text{ and } e^{ib\hat{q}}$ corresponding to translations in (continuous) position and and momentum space respectively obey a similar relation. 

Suppose now that $H$ is a product space, that is, $H= \tensor^m H_{d}$ where $H_d$ is a $d$-dimensional space and $N=d^m$. As a simple application of the basic relation \eqref{eq:basicReln} we show that the Fourier transform of product states in the computational basis are also product states and generalize a computationally useful formula. 
\begin{lem}
If the basis $\cali{B}_c$, the eigenvectors of $Z$, consists of $m$-fold product states then their Fourier transforms are also product states given by 
\[ \Omega(\ket{j}) =  (\sum_{r=0}^{m-1}\omega^{-jd^{m-1}r}\ket{r})\tensor\dotsb \tensor (\sum_{r=0}^{m-1}\omega^{-jdr}\ket{r})\tensor(\sum_{r=0}^{m-1}\omega^{-jr}\ket{r})\]
\end{lem}
\begin{proof}
Observe that there is an implicit ordering of the product states. Thus if $j=\sum_{r=0}^{m-1} d^rj_r$ is the representation of a positive integer $0\leq j \leq d^{m}-1$ then the state \(\ket{j} = \ket{j_{m-1}}\tensor \dotsb\tensor \ket{j_1}\tensor \ket{j_0}\equiv \ket{j_{m-1}}\dotsb\ket{j_1}\ket{j_0}\equiv \ket{j_{m-1}\dotsb j_0}\) where we suppress the tensor product symbol in the last two relations. Further, we write $\ket{0}$ for $\ket{0\dotsb 0}$

From the definition of QFT 
\[ 
\begin{split}
&\inpr{k_{m-1}\dotsb k_0}{\Omega}{j_{m-1}\dotsb j_0}= \inpr{0}{X^{-k}{\Omega}}{j_{m-1}\dotsb j_0} \\ & =\inpr{0}{\Omega\hconj{\Omega}X^{-k}{\Omega}}{j_{m-1}\dotsb j_0} =\inpr{0}{{\Omega}Z^{-k}}{j_{m-1}\dotsb j_0} \\
& =\omega^{-kj}\inpr{0}{\Omega}{j_{m-1}\dotsb j_0}= \frac{\omega^{-kj}}{\sqrt{N}}\\
\end{split}
\]
A direct computation shows that for the state
\[ \ket{\psi_j}= \frac{1}{\sqrt{N}}(\sum_{r=0}^{m-1}\omega^{-jd^{m-1}r}\ket{r})\tensor\dotsb \tensor (\sum_{r=0}^{m-1}\omega^{-jdr}\ket{r})\tensor(\sum_{r=0}^{m-1}\omega^{-jr}\ket{r})
\]
$\inp{k}{\psi_j}=\omega^{-kj}$. Since this is true for all $0\leq k \leq N-1$,   $\Omega{\ket{j}}=\ket{\psi_j}$. 
\end{proof}
\section{Distribution functions in quantum systems}\label{sec:quasprob}
One of the motivating factor's for  distribution functions in Wigner's work \cite{Wigner32} was the  construction of a quantum analogue of Liouville density in classical phase space. Following this approach suppose we want a  ``joint'' distribution function of the operators $X$ and $Z$. More precisely, we seek hermitian operators $\hat{x}$ and and $\hat{z}$ such that 
\beq\label{eq:gen2}
X = e^{i\genX} \text{ and } Z=e^{i\genZ}
\eeq
and then try to find distribution functions associated with the observables $\hat{x}$ and $\hat{z}$. 
We will do our computations in the computational basis $\cali{B}_c$ in which $Z$ is diagonal. It is easy to find $\genZ$. Thus 
\beq\label{eq:genZ}
\genZ= \frac{2\pi}{N}\begin{pmatrix} 0 & 0 & 0 & \dotsb & 0 \\ 0 & 1 & 0 & \dotsb & 0 \\ &&&&\\\vdots& & & \ddots &\vdots \\ 0& \dotsb & \dotsb & 0 & N-1\end{pmatrix}
\eeq
Of course $\genZ$ is only determined modulo $2\pi k$. 
From \eqref{eq:basicReln} and \eqref{eq:fourierMat} we have 
\beq\label{eq:genX}
(\genX)_{ij} = (\hconj{\Omega}\genZ\Omega)_{ij}= \sum_{k=0}^{N-1} k \omega^{k(i-j)} = \begin{cases} \pi(N-1)\text{ if } i=j \\ \frac{2\pi}{N}\frac{N-1}{\omega^{i-j}-1} \; i\neq j
\end{cases}
\eeq
These entries of the matrix $\genX$ imply that it is a hermitian {\em circulant} matrix. But a general linear combination $u\genX+v\genZ$ is not a circulant but a Toeplitz matrix. There are efficient algorithms for finding the eigenvalues and eigenvectors of such matrices. So in principle we can compute expressions like \(\ave{e^{i(u\genX+v\genZ)}}\) for real or integer $u,v$ using the standard diagonalization procedure. It is feasible to find analytic expressions, however, in low dimensions only. We will tackle the problem by different approach. Let us briefly review the continuous case first. 
\subsection{The Wigner distribution}
In the case of canonically conjugate variables like position and momentum a number of quasi-probability distributions are possible, each corresponding to a an operator ordering prescription. This is facilitated by the fundamental commutation relation 
\[ [\hat{q},\hat{p}] = i \hbar \]
between the position and momentum operators. Taking traces it is clear that such a relation is not possible in finite dimensions. So in finite dimension it is not clear how to prescribe ordering of operators. Moreover, there is some ambiguity in defining hermitian generators themselves. For example, for integers $a\text { and }b$, $\genX'= \genX+2\pi aI$ and $\genZ'= \genZ+2\pi bI$ are also infinitesimal generators for $X$ and $Z$ respectively but their linear combinations give rise to different set of unitary operators. The problem of non-uniqueness is essentially the same as the one that arises in defining roots and logarithms of complex numbers. Hence, we restrict to the principal branch of the logarithm as evident in the definition of $\genX$ and $\genZ$. 
\subsubsection{Continuous Wigner distribution}
The inversion formula of a characteristic function of classical probability is different for continuous and discrete probability distributions. 
In finite-dimensional quantum systems or more generally in the discrete part of the spectrum of a quantum observable we should use a formula analogous to that for discrete distributions \cite{Moyal}. But, this gives a quasi-probability distribution which may {\em not} have the desired properties \cite{Strat}. The problem seems to be rooted in the noncommutativity of quantum observables. 
The continuous Wigner distribution is defined by 
\beq\label{eq:contWigner}
W_c(x,z) \equiv \frac{1}{(2\pi)^2}\int\int\ave{e^{i(u\genX+v\genZ)}}e^{-i(ux+vz)}{\mathrm d}u {\mathrm d}v
\eeq
In this equation and the rest of the paper, unless the limits are explicitly stated, the real integrals are over the whole real line. Further we use the notation ${\bf r}= (x,y)^T$ for a real vector in 2 dimensions. 
The following theorem gives some of the important properties of the continuous Wigner distribution. First, we make the dependence on the state $\rho$ (mixed state, in general) explicit when necessary: $W_c(x,z:\rho)$. We give a simple proof of a well-known results in the appendix. 
\begin{thm}\label{thm:contWignerDist}
The function $W_c(x,z:\rho)$ is real. Moreover, it gives the correct marginal distributions.
\begin{equation}\label{eq:marginal}
\int_a^b\diffD z\int W_c(x,z:\rho){\mathrm d}x= \int_a^b \inpra{z}{\rho}  \diffD z
\end{equation}
where $\ket{z}$ are generalized eigenvectors of $\genZ$. We have a similar relation for the other marginal. We also have the following results on general marginal distribution. Let $R$ be an orthogonal matrix of order 2 representing a rotation. Let 
\[ {\bf r}' = \begin{pmatrix} x'\\ z'\end{pmatrix} = R {\bf r}\text{ with } R= \begin{pmatrix} \cos{\theta} & \sin{\theta} \\ -\sin{\theta} & \cos{\theta}\end{pmatrix}\]
Similarly we define orthogonal transformation in the ``noncommutative'' space. If 
\[ \begin{pmatrix} \genX'\\ \genZ'\end{pmatrix} = R \begin{pmatrix} \genX\\ \genZ\end{pmatrix} = 
\begin{pmatrix} \cos{\theta} \genX +\sin{\theta}\genZ \\ -\sin{\theta} \genX+\cos{\theta}\genZ\end{pmatrix}
\]
Then 
\beq\label{eq:Radon_marginal}
\int_a^b\diffD z'\int W_c(x,z:\rho){\mathrm d}x'= \int_a^b\diffD z'\inpra{z'}{\rho}
\eeq
where $\ket{z'}$ are generalized eigenvectors of $\genZ'$ and the variables $x\text{ and }z$ are considered as functions of $x',z'$. A similar result holds for the conjugate observable $\genX$. 
\end{thm}
The continuous version of distribution function of discrete observables is problematic. First, we want the marginals to resemble classical discrete probability distributions so we have delta functions at the isolated points. To justify the later we have to integrate over some domain of the continuous variable and this causes some problems interpreting these as probability distributions. Some authors have attempted to tackle these problems by focusing on  continuous families of discrete observables like spin direction. These approaches seem somewhat unnatural to us. Discrete distributions must be characterized by a discrete measure (e.g.\ the counting measure) and thus the integrals must be replaced by sums. In particular, for finite systems we must have finite sums. This is the avenue we will explore in this paper. Finally, let us mention an important point. The Wigner distribution and some other related probability and quasi-probability distributions are sometimes interpreted as {\em joint} probability distributions of incompatible observables. Clearly, any measurement of such distribution must give unsharp values of these observables, consistent with the uncertainty principle. The Arthrus-Kelley scheme \cite{Arthurs65, Braunstein91} is an example. For discrete observables the concept of joint distribution of noncommuting observables is difficult even for fuzzy measurements. 
\section{Discrete Quasiprobability Distributions}\label{sec:di_quasiprob}
The Wigner and other distribution functions are an alternative to the  density matrix formulation of quantum theory and are given by  distribution function $W({\bf y}:\rho)$ with ${\bf y}$ representing classical parameters. Expectation values of any quantum mechanical quantity that can be computed in a given state $\rho$ can be computed from $W({\bf y}:\rho)$. Hence we have a correspondence between quantum observables and ``classical'' observables along with an ordering prescription. Since, the density matrix provides a maximal description of a quantum system so does $W({\bf y}:\rho)$. We thus have an alternative semiclassical picture. In some situations the latter may be easier to determine experimentally. In any case, such quasiprobability distributions provide a useful tool for visualization. 

\subsection{Properties of Distribution Functions}\label{sec:propDist}
Let us make precise the requirements we impose on distribution functions. Let $W({\bf y}: \rho)$ be a distribution function associated with a quantum state $\rho$ of a quantum system $S$ and ${\bf y}$ is a real vector representing a finite set of ``classical'' parameters. Let $H$ be the system Hilbert space and $\cali{S}(H)$ the set of states, that is, the {\em convex} set of normalized positive trace-class operators. 
\be 
\item[R1.]
$W({\bf y}: \rho)$ is a continuous real function on $\cali{S}(H)$ that preserves convex combinations: if $\rho_1,\rho_2\in \cali{S}(H)$ and $0\leq s \leq 1$ then 
\[W({\bf y}: s\rho_1+(1-s)\rho_2)= sW({\bf y}: \rho_1)+(1-s)W({\bf y}: \rho_2)\] 
It is nondegenerate in the sense that at no point in phase space $W({\bf y}:\rho)$ is identically 0 for all $\rho$. 
\item[R2.]
For two states $\rho$ and $\rho'$ 
\beq
 \tr(\rho\rho')= K\int W({\bf y}:\rho)W({\bf y}:\rho')\diffD {\bf y}
 \eeq
Part of the above requirement is that we define the appropriate measure $\diffD {\bf y}$ which also fixes the constant $K$. The constant $K=2\pi\hbar$ for continuous systems and $K=N$ for finite system of dimension $N$. This constant equals the volume of a ``phase space cell''. With respect to this measure we also demand normalization condition 
\[ \int W({\bf y}: \rho) \diffD {\bf y}=1\]
Note that this is a nontrivial requirement as this implies that the left side of the above equation must be independent of the quantum state. 
\commentout{
\item[R3.]
The distribution function $W(x,z:\rho)$ gives the correct marginals. More, precisely
\beq
\sum_x W(x,z:\rho)= \delta_{xj} \tr(\pj{j}\rho) \text{ and } \sum_z W(x,z:\rho)= \delta_{zj} \tr(\pj{\tilde{j}}\rho)
\eeq
where $\pj{\tilde{j}}$ (resp. $\pj{\tilde{j}}$) are eigenvectors of $z$ (resp. $x$) with eigenvalue $2\pi j/N$. }
\item[R3.]
For any observable $A$ on $S$ there is a  real function $\tilde{A}({\bf y})$ such that the expectation value (in state $\rho$)
\beq \label{eq:normalization}
\ave{A}= \tr(\rho A)= \int W({\bf y}:\rho) \tilde{A}({\bf y})\diffD {\bf y}
\eeq
\ee

The first requirement is that $W$ must be real. If we try to impose nonnegativity however it becomes too stringent. As $W({\bf y}: \rho)$ is convex linear on states in finite dimensional spaces it has a unique extension to a liner functional (for fixed {\bf y}) on $\cali{K}(H)$ the set of bounded hermitian operators (observables). In infinite dimensions we need some delicate continuity arguments. Henceforth, we will assume linearity of  $W({\bf y}: \rho)$. In these specifications for distribution function we have not mentioned marginals. We will discuss them shortly. What are the characteristics of the parametric vector ${\bf y}$? If it is to be somehow identified with generators of classical observables its dimension must be related to degrees of freedom. The third item in the list gives a clue. A physical system, whether classical or quantum, is completely characterized by the set of observables $\cali{O}$. Often $\cali{O}$ has more structure, in particular, it is an algebra. The difference between quantum and classical algebra of observables is that the former is {\em noncommutative}. These algebras have minimal sets of generators. For example, the observable algebra of a classical system with $N$ degrees of freedom is generated by generalized coordinates $\{ q_i:i=1,\dotsc,N\}$ and the conjugate momenta $\{p_i:i=1,\dotsc,N\}$. The corresponding quantum algebra is also generated by the {\em operators} $\hat{q}_i\text{ and } \hat{p}_i$ which do not commute. In the finite dimensional case we have no classical analogue. But we will be guided by this example. We have already discussed the close analogy between the finite-dimensional unitary operators $X,Z$ and the continuous operator $e^{i\hat{p}}, e^{i\hat{q}}$. We show next that $\hat{x}\text{ and }\hat{z}$ are actually generators of the {\em complex} algebra of observables in the appropriate dimension. 
\begin{propn}
Let the dimension of the system Hilbert space be $N$ and $\genX$ and $\genZ$ be as given in \eqref{eq:genX}  and  \eqref{eq:genZ} respectively. The completion of the complex algebra generated by $\genX$ and $\genZ$ equals $M_n(\complex)$, the algebra of complex matrices of order $N$. 
\end{propn}
\begin{proof}
The completion of algebra means that we include the limits of convergent sequences. In particular, $X=e^{i\genX}$ and $Z=e^{i\genZ}$ are in the completion. We show that $X$ and $Z$ generate $M_n(\complex)$. Let $\omega=e^{2\pi i/N}$ and $Z_k=\omega^{-k}Z$. It is easy to see that 
\[ (I+Z_k+Z_k^2+\dotsb+Z_k^{N-1})/N= D_k\]
where $D_k(ij)=\delta_{ij}\delta_{jk}$ is the diagonal matrix with 1 in the $k$th row (and column) and 0's everywhere. We also see that $X^jD_k=E_{j+k,k}$ where $E_{ij}$ are the elementary matrices with 1 in the $ij$th place and 0's everywhere else. Note that $j+k$ is to be considered $\mod{N}$. Thus every elementary matrix is generated by $\genX$ and $\genZ$. As the elementary matrices constitute a basis for $M_n(\complex)$ the proof is complete. 
\end{proof}
We mention that the assertion of the proposition was essentially proved by J. Schwinger \cite{Schwinger70} in different way. The continuous quasiprobability distribution functions can be written as 
\beq\label{eq:WignerGenC}
W({\bf q},{\bf p})=\int f({\bf u},{\bf v})\ave{e^{i\hat{\bf q}\cdot {\bf u}}e^{i\hat{\bf p}\cdot{\bf v}}}e^{-i({\bf u}\cdot{\bf q}+{\bf v}\cdot{\bf p})}\diffD {\bf u}\diffD {\bf v} 
\eeq
Here $f$ is a scalar or $c$-number function which is usually interpreted as an operator ordering prescription. The Wigner distribution function is a special case corresponding to Weyl ordering. All this is possible because of the simple commutation properties of the observables $\hat{q}_i\text{ and } \hat{p}_i$. We have observed that the unitary operators $X$ and $Z$ have multiplicative relations very similar to $e^{i\hat{p}}$ and $e^{i\hat{ q}}$ (see equation \eqref{eq:commut}). This analogy extends further
\beq
Z^aX^b= \omega^{ab} X^bZ^a;\quad e^{ia\hat{q}}e^{ib\hat{p}}= e^{iab}e^{ib\hat{p}}e^{ia\hat{q}}
\eeq
provided $a$ and $b$ are integers. Of course, the second formula is valid for all real $a$ and $b$ but the first fails if {\em both} are non-integer. This provides another reason to construct a discrete version of distributions functions. Henceforth we will restrict to finite dimensional spaces mostly. Since the operator $\hat{x}$ and $\hat{z}$ can be used as generators we will assume the ``phase space'' spanned by ${\bf y}=(x,z)$ is 2-dimensional. Now we can state the marginal conditions corresponding to the ``axes''. 
\bi
\item[R4.] The quasiprobability distribution function $W(x,z)$ have marginal distributions that coincides with probability distributions of the {\em quantum} observables $\hat{x}$ and $\hat{z}$:  
\beq
\sum_x W(x,z:\rho)= \delta_{xj} \tr(\pj{j}\rho) \text{ and } \sum_z W(x,z:\rho)= \delta_{zj} \tr(\pj{\tilde{j}}\rho)
\eeq
where $\pj{\tilde{j}}$ (resp. $\pj{\tilde{j}}$) are eigenvectors of $z$ (resp. $x$) with eigenvalue $2\pi j/N$.
\ei
We seek a finite distribution function similar to the form \eqref{eq:WignerGenC} above. For a state $\rho$ in a finite quantum system of dimension $N$ define
\beq \label{eq:WignerGenD}
\begin{split}
W(x,z:\rho) &= \sum_{m,n=0}^{N-1} f(m,n)\ave{X^mZ^n} \omega^{-(mx+nz)}\\
& \text{ with } \omega = e^{2\pi i/N} \text{ and } 0\leq j,k\leq N-1 \text{ integers }\\
\end{split}
\eeq
Call the functions $f$ in the above expression {\em ordering} functions. To compute the expectation values we need the following matrix elements in computational basis. 
\beq \label{eq:matEl}
\inpr{k}{X^mZ^n}{j}= \begin{cases}
\delta_{m,k-j}\omega^{jn} \text{ if } k\geq j\\
\delta_{m,N+k-j} \omega^{jn}\text{ if } k <j
\end{cases}
\eeq
To see the implications of the reality condition R1 it is sufficient to verify it for pure states. Hence for $\rho=\pj{\alpha}$
\[
\begin{split}
\overline{W(x,z:\rho)}&= \sum_{m,n=0}^{N-1} \overline{f(m,n)} \inpr{\alpha}{Z^{-n}X^{-m}}{\alpha}\omega^{mx+nz}\\
&= \sum_{m,n=0}^{N-1} \overline{f(m,n)} \omega^{mn}\inpr{\alpha}{X^{-m}Z^{-n}}{\alpha}\omega^{mx+nz}\\
&= \sum_{m,n=1}^{N} \overline{f(N-m,N-n)} \omega^{mn}\inpr{\alpha}{X^{m}Z^{n}}{\alpha}\omega^{-(mx+nz)}\\
&= {W(x,z:\rho)}\\
\end{split}
\]
In the last step we use $X^N=Z^N=I$. Since this must hold for all state vectors $\alpha$ we have 
\beq\label{eq:R1}
\begin{split}
& \overline{f(N-m,0)}= f(m,0), \quad \overline{f(0,N-n)}=f(0,n)\text{ and } \\
&\overline{f(N-m,N-n)} \omega^{mn}= f(m,n),\; 1<m,n <N
\end{split}
\eeq
We will see later that the condition of nondegeneracy is automatically satisfied. 
Next we consider R2. Let $\rho = \sum \rho_{jk}\pj{j}{k}$ and $\rho' = \sum \rho'_{jk}\pj{j}{k}$. Then using \eqref{eq:matEl}  
\beq\label{eq:TrComp_base}
\begin{split}
&\sum_{xz}W(x,z:\rho)W(x,z:\rho') \\
&=\sum_{xz}\sum_{\substack{jmn\\j'm'n'}}f(m,n)f(m',n')\rho_{j,j+m} \rho_{j',j'+m'}\omega^{jn+j'n'} \omega^{-((m+m')x+(n+n')z)}\\
&= {N^2}\sum_{jmn}\sum_{j'}f(m,n)f(N-m,N-n)\rho_{j,j+m} \rho_{j',j'+N-m}\omega^{(j-j')n}\\
&= {N^2}\sum_{jmn}\sum_{j'}f(m,n)\overline{f(m,n)}\rho_{j,j+m} \rho_{j',j'+N-m}\omega^{(j-j'+m)n}\\
\end{split}
\eeq
where we have used \eqref{eq:R1} in the last step. According to R2 this should be equal to $\tr(\rho\rho')/N=(\sum_{jk} \rho_{jk}\rho'_{kj})/N$ for all choices of density matrices $\rho$ and $\rho'$. This is possible if $|f(m,n)|^2$ is a constant independent of $m$ and $n$. A straightforward computation yields $|f(m,n)|=1/N^2 $. Setting $f(m,n)=g(m,n)/N^2$ we may write $g(m,n)=\omega^{\beta(m,n)}$. We now prove existence and properties of distribution functions satisfying the conditions R1-R4. 
\begin{thm}\label{thm:existWigner1}
For every density matrix $\rho$ in an $N$-dimensional Hilbert space and $\omega=e^{2\pi i/N}$ let 
\[W(x,z:\rho)= \sum_{m,n=0}^{N-1} f(m,n)\ave{X^mZ^n} \omega^{-(mx+nz)}\]
Then there exist functions $f(m,n)$ such that the corresponding $W$ satisfies R1,R2 and R4. Moreover, for any $W$ satisfying these conditions there are unique hermitian operators $\hat{a}(x,z)$ such that following hold. 
\begin{subequations}\label{eq:propWigner}
\begin{align}
& W(x,z:\rho)=\tr(\rho \hat{a}(x,z)) \text{ and } \rho= N\sum_{xz} W(x,z:\rho)\hat{a}(x,z)\label{eq:propWigner1}\\ 
& \tr(\hat{a}(x,z)\hat{a}(x',z'))=\frac{1}{N}\delta_{xx'}\delta_{zz'} \text{ and } \sum_{xz} \hat{a}(x,z) = I\label{eq:propWigner2}
\end{align}
\end{subequations}
Given a hermitian operator $T$ let $t(x,z)= \tr(T\hat{a}(x,z))$ then 
\beq \label{eq:R4}
\ave{T}= \sum_{xz} W(x,z:\rho) t(x,z) 
\eeq
Thus R3 is also satisfied. 
\end{thm}
\begin{proof}
We have observed that functions $f(m,n) = \omega^{\beta(m,n)}/N^2$ satisfying the relations \eqref{eq:R1} provide a distribution function $W(x,z:\rho)$ that satisfies conditions R1 and R2. To see the implications of condition R4 on marginals we observe that 
\beq
\begin{split}
&\sum_x W(x,z:\rho)  = \sum_{mn} f(m,n)\ave{X^mZ^n} \omega^{-nz}\sum_x \omega^{-mx} \\
& = \sum_{mn} f(m,n)\ave{X^mZ^n} \omega^{-nz}\delta_{m0}= N\sum_{n} f(0,n)\ave{Z^n} \omega^{-nz}\\
& =  N\sum_{jkn} f(0,n) \rho_{jk} \inpr{k}{Z^n}{j}\omega^{-nz}
= N\sum_{jn} f(0,n) \rho_{jj} \omega^{(j-z)n}
\end{split}
\eeq
For the last expression to be equal to $\tr(\rho\pj{z})$, the probability of finding the system in an eigenstate of $\hat{z}$ with eigenvalue $2\pi z/N$, we must have $f(0,n)=1/N^2\text{ for all }n$. Computing the trace in the Fourier transformed basis $\ket{\tilde{j}}=\Omega\ket{j}$ we conclude that the second condition in R3 yields $f(m,0)=1/N^2$. Assuming $\beta(m,n)$ to be a real polynomial in $m\text{ and } n$ we conclude that it must be of the form $\beta(m,n)=mn\alpha(m,n)$. More generally, we may take $\beta(m,n)= \gamma(m,n)+mn\alpha(m,n)$ with $\gamma(0,n)=\gamma(m,0)=0$. With this choice of $\beta(m,n)$ the first set of equations in \eqref{eq:R1} are satisfied. The second set gives the following requirement on the function $\alpha$. 
\beq\label{eq:basicEqOrder}
\begin{split}
&mn(\alpha(m,n)-1)+(N-m)(N-n)\alpha(N-m,N-n)+\\
&\gamma(m,n)+\gamma(N-m,N-m)= 0 \mod{N}
\end{split}
\eeq
Note that we do not require $\alpha$ or $\gamma$ to be integer-valued or symmetric. There exist (real) functions satisfying equation \eqref{eq:basicEqOrder} for all $0\leq m,n\leq N-1$. Simple solutions to these equations are given below. 
\beq\label{eq:order1}
f_0(m,n) = \begin{cases}  \frac{\omega^{mn(N+1)/2}}{N^2} \text{ if } N \text{ odd }\\
\nu_{mn}\frac{\omega^{mn/2}}{N^2}, \; N \text{ even }
\end{cases}
\eeq
where $\nu_{mn}$ satisfies 
\beq\label{eq:extraPhase_even}
|\nu_{mn}|=1 \text{ and }\conj{\nu_{N-m,M-m}}=(-1)^{m+n}\nu_{mn}
\eeq
\commentout{
\begin{cases}
=\frac{\omega^{mn/2}}{N^2}, \; N \text{ even and }m+n\text{ even or }mn=0\\
=i\frac{\omega^{mn/2}}{N^2}, \; N \text{ even and }m+n\text{ odd and }mn>0
\end{cases}
\end{cases}
\eeq 
}
A particular choice of $\nu$ satisfying \eqref{eq:extraPhase_even} is 
\beq\label{eq:evenSpecial}
\nu_{mn}= \omega^{(1-\delta_{m0})(1-\delta_{n0})(m+n)^2N/4}
\eeq
Other choices of $\nu_{mn}$ will be given later when we impose more conditions on the distribution functions.
Finally, suppose the functions $f$ in the definition of $W(x,z:\rho)$ satisfy the reality conditions \eqref{eq:R1} and the marginal condition $f(m,0)=f(0,n)=1/N^2$. Then it is easy to see that for the incoherent state $I/N$ 
\beq
W(x,z:\frac{I}{N})= \frac{1}{N^2}
\eeq
The distribution function is {\em nondegenerate} at each point in phase space. 
This proves the existence of solutions to equations \eqref{eq:basicEqOrder} and hence quasiprobability distributions satisfying R1,R2 and R4 in all finite dimensions. 

Observe that the map $\Xi(x,z): \rho\rightarrow W(x,z:\rho)$ is real and can be uniquely extended to a  linear map on all hermitian operators. That is, $\Xi$ is a linear functional on $K_H$, the linear space of hermitian operators on the system Hilbert space $H$. $K_H$ is a real Hilbert space with respect to the scalar product $<A,B>=\tr(AB),\; A,B\in K_H$. Since $\Xi(x,z)$ is nondegenerate at each point there exists a unique nonzero $\hat{a}(x,z)\in K_H$ such that $W(x,z:\rho)=<\hat{a}(x,z),\rho>=\tr(\hat{a}(x,z)\rho)$. So the first of the equations in \eqref{eq:propWigner1} holds. The condition R2 and \eqref{eq:propWigner1} together imply 
\[
\begin{split}
&<\rho,\rho'>=\tr(\rho\rho') = \sum_{xz} W(x,z:\rho)W(x,z:\rho') \\
&= \sum_{xz}W(x,z:\rho) \tr(\hat{a}(x,z)\rho')=\tr((\sum_{xz}W(x,z:\rho) \hat{a}(x,z))\rho')\\
&=<\sum_{xz}W(x,z:\rho) \hat{a}(x,z),\rho'>\\
\end{split}
\]
Since this is true for all positive definite operators $\rho'$ with trace 1 we conclude that the second of the equation in \eqref{eq:propWigner1} must hold. Now using this expansion of $\rho$ in the operators $\hat{a}(x,z)$ and that fact R2 again we conclude that the equations \eqref{eq:propWigner2} hold. 
Finally, to prove that condition R3 also holds observe that any hermitian operator $T$ can be written in the form $T= b_1\rho_1-b_2\rho_2$, with $b_1,b_2>0$ and $\rho_1,\rho_2$ density matrices. Then,
\[
\begin{split}
&\ave{T}=\tr(\rho T)=<\rho,T>=b_1<\rho,\rho_1>-b_2<\rho,\rho_2>\\
&= b_1\tr((\sum_{xz}W(x,z:\rho)\hat{a}(x,z))\rho_1)- b_2\tr((\sum_{xz}W(x,z:\rho)\hat{a}(x,z))\rho_2) \\
&= \sum_{xz}W(x,z:\rho)\tr(\hat{a}(x,z)(b_1\rho_1-b_2\rho_2))=\sum_{xz}W(x,z:\rho)\tr(\hat{a}(x,z)T)\\
&= \sum_{xz}W(x,z:\rho)t(x,z)\\
\end{split}
\]
\end{proof}
The distribution functions corresponding to $f_0$ will be called (finite) Wigner functions. In case of odd dimension there is one such function. But in even dimensions we have to be more careful in our choices. 
\subsection{Explicit formulas}
The mere existence of ``orthonormal'' hermitian operators like $\hat{a}(x,z)$ which span the (real) space of observables  is simply a statement about the existence of orthonormal bases in any Hilbert space. Two characteristics  distinguish $\hat{a}(x,z)$: first the marginal distributions associated with them (R4) and second the way they were derived via the quantum Fourier transform. Our next task is to find explicit forms for these operators. Let 
\beq
W(x,z:\rho,f)= \sum_{m,n} f(m,n) \ave{X^mZ^n} \omega^{-(mx+nz)}
\eeq
be a quasiprobability distribution satisfying R1-R4. We have indicated explicit dependence on the ordering function $f$. From this it follows that the phase-point operators are given by 
\beq\label{eq:phaseOp}
\hat{a}(x,z:f) = \sum_{m,n} f(m,n) X^mZ^n \omega^{-(mx+nz)}
\eeq
The fact that $\hat{a}(x,z:f)$ form an orthonormal operator basis can be verified directly. The name ``phase-point operator'' derives from the fact that $(x,z)$ may be considered as a ``point'' in a {\em finite} phase space. The quasiprobability distribution $W(x,z:\rho,f)$ are simply the coefficients in the expansion of $\rho$ in the basis  $\{\hat{a}(x,z:f)\}$. 
We will compute these operators in the ``computational'' basis $\{ \ket{j}=\ket{j\mod N}\}$, that is, the eigenbasis of the operator $Z$. Then \( X^m= \sum_j \prj{j+m}{j}\text { and } X^mZ^n= \sum_j \omega^{jn} \prj{j+m}{j}\). A straightforward calculation then gives
\beq \label{eq:basicExp}
\hat{a}(x,z:f)_{kl} \equiv \inpr{k}{\hat{a}(x,z:f)}{l}= \omega^{-(k-l)x}\sum_n  f(k-l,n)\omega^{n(l-z)}
\eeq
In particular, the diagonal terms are easy. 
\beq
\hat{a}(x,z:f)_{kk}= \delta_{k,z} /N
\eeq
Now using the formulas \eqref{eq:order1} for $f(m,n)$ in the formula we get the following two cases for $N$. First for $N$ odd, 
\beq\label{eq:PhaseOp_odd}
\hat{a}(x,z:f_0)_{kl} =\frac{ \omega^{-(k-l)x}}{N^2}\sum_n  \omega^{n(k-l)(N+1)/2}\omega^{n(l-z)}
= \frac{ \omega^{-(k-l)x}\delta_{k+l,2z}}{N}.\\
\eeq
Apart from ordering and normalization these are precisely the phase-point operators found in \cite{Wootters87} for prime dimensions. Note that we do not require the dimension $N$ to be prime. If $N=2r$ is even the calculation is a bit more involved as the corresponding expression for $f_0(m,n)$ in \eqref{eq:order1} is not ``homogeneous''. We now have 
\beq\label{eq:PhaseOp_evenGen}
\hat{a}(x,z:f_0)_{kl} \equiv \tr(\prj{l}{k}\hat{a}(x,z:f))= \omega^{-(k-l)x}\sum_n  \nu_{mn}\omega^{n(k+l-z)/2}
\eeq
Evaluating these sums is not difficult but one has to be careful about the signs. For the choice of $\nu_{mn}$ given in \eqref{eq:evenSpecial} we get  
\beq\label{eq:PhaseOp_even}
\hat{a}(x,z)_{kl}=\begin{cases}
\frac{\omega^{-(k-l)x}(1\pm i)}{2N}\delta_{k+l,2z}\quad k-l\text{ even}\\
\frac{\omega^{-(k-l)x}\cot{(\pi(k+l-2z)/N)}\pm i\csc{(\pi(k+l-2z)/N)}}{N^2} \quad k-l\text{ odd}
\end{cases}
\eeq
So we see that the quasiprobability functions given above are much more complicated in even dimension. More importantly, the phase-point operators given by \eqref{eq:PhaseOp_odd} is more {\em sparse} than the one \eqref{eq:PhaseOp_even} for even dimension. This, in turn, implies that in general quasiprobability distributions are sparser in the odd dimension and ``computationally simpler''. Let us illustrate with an example. 

Suppose a quantum circuit or protocol is supposed to produce a state $\ket{b}$ in the computational basis. Because of noise and imperfections we actually get a state (possibly mixed) which lies in the state space corresponding to the subspace $K$ spanned by $\{\ket{b\pm i}: i\leq a\}$. From the formulas \eqref{eq:PhaseOp_odd} and \eqref{eq:PhaseOp_even} it is easy to see that the number of nonzero entries $W(x,z)$ in the odd case is $O(a)$ and in the even case it is $O(a^2)$. From the duality between $X$ and $Z$ this is also true if the computational basis is replaced by its Fourier transform. Since finding quasiprobability distribution equivalent to determining the state does it mean that odd dimensions are tomographically ``better''? That we should look at qutrits too? 
\section{Heisenberg groups}\label{sec:Heis}
In this section we turn to our  main theme: the Heisenberg groups and their close connections with Fourier transforms and distribution functions (see \cite{Folland89} for this connection in the continuous case). There are families of continuous and discrete Heisenberg groups. Although our primary focus will be on the discrete Heisenberg groups we first take a look at the continuous Wigner function from a different perspective. We start with the (continuous) $n$-dimensional Heisenberg group ${\mathbf H}^n$ whose group manifold is $\real^{2n+1}$. Using vector notation we write the elements as $(\vect{p},\vect{q},t)$ where $\vect{p}\text{ and }\vect{q}$ are vectors in $\real^n$ and $t$ is a real number. The reader can easily recognize the ``phase space'' behind this notation. The group multiplication is defined by 
\[ (\vect{p},\vect{q},t)(\vect{p}',\vect{q}',t')=(\vect{p}+\vect{p}',\vect{q}+\vect{q}',t+t'+(\vect{p}\cdot\vect{q}'-\vect{q}\cdot\vect{p}')/2)\]
where the $\cdot$ denotes the usual scalar product. The symplectic structure is apparent in the above definition. By changing the parametrization of the group \((p,q,t)\rightarrow (p,q,t-pq/2)=(p',q',t')\) we get the multiplication law of the (polar) Heisenberg group \cite{Folland89}. 
\beq
(p'_1,q'_1,t'_1)(p'_2,q'_2,t'_2)=(p'_1+p'_2,q'_1+q'_2,t'_1+t'_2+p'_1q'_2)
\eeq
Note that the element $(0,0,t)$ is in the centre of the group. Let us restrict to $n=1$ for simplicity. The Lie group ${\mathbf H}^1$ is generated by the Lie algebra $\mathfrak{h}_1$ with generators $\{p,q,\lambda\}$ with brackets $[p,q]=\lambda,\text{ and } [\lambda,p]=[\lambda,q]=0$. One constructs the Poisson structure on the dual space $\mathfrak{h}_1^*$ in a natural way. The Heisenberg group plays a fundamental role in quantum mechanics. The Stone-von Neumann theorem  asserts that the standard representation of position and momentum are essentially unique. In other words, the Schroedinger picture 
\[ (p,q,t) \rightarrow \gamma(p,q,t)\equiv e^{2\pi it}e^{2\pi i(p\hat{p}+q\hat{q})} \]
with $\hat{q}\psi(q')=q'\psi(q')$ and $\hat{p}\psi(q)=-i\frac{\partial \psi(q)}{\partial x}$ is the unique representation of Heisenberg group under some conditions of continuity. Here $\psi$ is the wave function in one dimension. Mathematically, it lives in the space $H=L^2(\real)$ of complex square integrable functions (we ignore the technical difficulties arising due to the unboundedness of the operators). Since, the elements $(0,0,t)$ are in the centre it is often sufficient to consider only elements of the form $\gamma(p,q)=\gamma(p,q,0)= e^{2\pi i(p\hat{p}+q\hat{q})}$.  This is the reduced Heisenberg group. Let $\{\psi_{\alpha}(x) \}$ be a basis in $H$. The matrix elements in this basis are given by 
\[ 
\begin{split}
 &V_{\alpha\alpha'}(p,q)= \inpr{\psi_\alpha}{\gamma(p,q)}{\psi_{\alpha'}}= \inpr{\psi_\alpha}{e^{2\pi i(p\hat{p}+q\hat{q})}}{\psi_{\alpha'}}\\
 &=\int \conj{\psi_\alpha(u)}{e^{2\pi i(p\hat{p}+q\hat{q})}}\psi_{\alpha'}(v)\diffD u\diffD v
\end{split} 
\]
These are precisely matrix elements of the Fourier transform of the phase-point operators in the continuous case. In particular, $V_{\alpha\alpha}(0,q)$ yields Fourier transforms of the position probability density corresponding to the state $\psi_\alpha$. Similarly, using the momentum representation we get the other marginal for $V_{\alpha\alpha'}(p,0)$. Since the basis was arbitrary we conclude that the Wigner function 
\[ W(p,q)= \int \ave{\gamma(u,v)}e^{-2\pi i(pu+qv)}\diffD u\diffD v\]
when integrated over the strip between $q=c_1$ and $q=c_2$ gives the probability of the particle in (pure) state $\psi$ to have its {\em position} between $c_1$ and $c_2$. Explicitly, 
\[\int_{c_1}^{c_2}\diffD q\int_{-\infty}^{\infty}W(p,q:\ket{\psi})\diffD p\]
yields the probability that the position observable has value between $c_1$ and $c_2$ and similarly for the momentum observable. This is easily seen by expanding $\ket{\psi}$ in the position basis. What do we get if we integrate over an arbitrary strip, not necessarily parallel to the $p$ or $q$ axes, say the lines $ap+bq=c_1$ and $ap+bq=c_2$? The answer is well-known and is discussed in \cite{Wootters87} and \cite{Hillery84}. But we look at it from a different perspective. First, put $ap+bq=p'$. This defines a family of parallel lines $p'=c$ in the $p$-$q$ plane. Another line $cp+dq=q'$ does {\em not} belong to this family if and only if $ad-bc\neq 0$. Thus, the matrix $\zeta=\begin{pmatrix} a & b\\c & d\end{pmatrix}$ is invertible and defines a change of coordinate in the phase plane. Then the form $up+vq=u'p'+v'q'$ where $(u', v')=(u,v)\zeta$. This in turn defines a transformation on the Lie algebra generated by $\hat{p},\hat{q}$:
\[\gamma(u,v)= {e^{2\pi i(u\hat{p}+v\hat{q})}}={e^{2\pi i(u'\hat{p}'+v\hat{q}')}},\quad 
\begin{pmatrix}
\hat{p}' \\ \hat{q}'
\end{pmatrix}
=\zeta \begin{pmatrix}\hat{p} \\ \hat{q}\end{pmatrix}
\]
If the transformation $\hat{p}\rightarrow \hat{p}',\hat{q}\rightarrow \hat{q}'$ were an {\em automorphism} then the $\hat{p}'$ and $\hat{q}'$ have the same commutation relation as $\hat{p}$ and $\hat{q}$. This will happen if and only if $\det \zeta=ad-bc=1$. But then if we change the variable of integration to $p',q'$ the measure remains unchanged ($|\det \zeta|=1$). We can now carry over the argument from the case of axes marginals and conclude that integration of $W(p,q:\ket{\psi})$ over a strip between $ap+bq=c_1$ and $ap+bq=c_2$ gives the probability that the observable $\hat{p}'=a\hat{p}+b\hat{q}$ will have value lying between $c_1$ and $c_2$. Let us observe that $\zeta\in SL(2,\real)=Sp(1,\real)$ where $SL(n,\real)$ is the group of $n\times n$ real matrices with determinant 1 and $Sp(n,\real)$ is the real symplectic group of order $n$. In general for  $Sp(n,\real)$ is a subgroup of the automorphism group of ${\mathbf H}^n$ and is different from $SL(2n,\real)$. 

Now we turn to the discrete Heisenberg group ${\bf H}$. We define a {\em presentation} of the group in terms of generators and defining relations \cite{Rotman}. $\heis{}$ is generated by $\{{\bf x},{\bf z}, \gamma\}$. The defining relations are  
\beq
{\bf z}{\bf x}= \gamma{\bf x}{\bf z},\;\gamma{\bf x} ={\bf x}\gamma \text{ and } \gamma{\bf z} = {\bf z}\gamma 
\eeq
The advantage of this approach is that any map $\phi$ from the generators of a group $\heis{}$ to another group $K$ which satisfies the same defining relations as above can be uniquely extended to a group {\em homomorphisms} $\heis{}\rightarrow K$.
A simple realization of the group over integers is given by the set ${\mathbf Z}^3$. The multiplication is defined by 
\[
(j_1,k_1,t_1)(j_2,k_2,t_2)=(j_1+j_2,k_1+k_2,t_1+t_2+j_1k_2)
\] 
The generators are ${\bf z}= (1,0,0),{\bf x}= (0,1,0),\text{ and } \gamma=(0,0,1)$. If we specialize to ${\mathbf Z}_N$, the integers modulo $N$ we get corresponding Heisenberg group ${\bf H}_N$ with generators $X,Z,\text{ and }\gamma$ and the relations
\beq
X^N=Z^N=\gamma^N=e\text{ (identity)},\gamma X=X\gamma,\gamma Z=Z\gamma\text{ and } ZX=\gamma XZ
\eeq

Since $\heis{N}$ is a finite group its finite-dimensional representations are unitary and completely reducible. Let $\phi$ be a representation of $\heis{}$ or $\heis{N}$ on a vector space $V$ of finite dimension. We say that the central element $\gamma$ acts {\em maximally} if the order of $\phi(\gamma)$ is $\dim(V)$. The following theorem characterizes representation of $\heis{}\;(\heis{N})$ and their relation to QFT.  
\begin{thm}\label{thm:connectHeisWigner}
Let $\phi$ be a (unitary) irreducible representation of $\heis{N}$ on a finite-dimensional space $V$. Let $\tau$ be the automorphism of $\heis{N}$ (and $\heis{}$) given by \(X\rightarrow Z, Z\rightarrow X\text{ and }\gamma\rightarrow \gamma^{-1}\) and $\phi'$ the representation defined by $\phi'(g)\equiv \phi(\tau g)$. Then the following statements are true. 
\be
\item
$\phi(\gamma)=\omega$, a primitive $N$th root of 1 and the eigenvalues of $\phi(Z)$ and $\phi(X)$ are $\{\omega^k:0\leq N-1\}$. $\gamma$ acts maximally if and only if $\phi$ is faithful (one-to-one). 
\item 
$\phi$ and $\phi'$ are unitarily equivalent: \(\phi'=\Omega\phi\hconj{\Omega}\) and $\Omega$ is the quantum Fourier operator. 
\item
Any unitary irreducible representation $\psi$ of the full discrete Heisenberg group $\heis{}$ in $V$ in which the order of $\gamma$, $o(\gamma)=\dim{(V)}=K$ is equivalent to an irreducible faithful representation of $\heis{K}$.  
\ee
\end{thm}
\begin{proof}
Assume first that $\gamma$ acts maximally. Since $\phi$ is irreducible and $\gamma$ is in the centre it must act as a constant (Schur's lemma). As order of $\gamma$ is $N$, $\gamma=\omega I$ where $\omega$ is a primitive $N$th root of unity. Since $\heis{N}$ is finite we may assume the representations to be unitary. Let $\alpha$ be an eigenvector of $\phi(Z)$ with eigenvalue $c$. As $Z^N=e$, $c$ must be an $N$th root of 1. Consider the set $S=\{\alpha, \phi(X)\alpha, \dotsc, \phi(X^{N-1})\alpha\}$. As \[\phi(Z)\phi(X^k)\alpha=\phi(\gamma^k)\phi(X^k)\phi()\alpha=c\omega^k\phi(X^k)\alpha\] $\phi(X^k)\alpha= \phi(X)^k\alpha,\; k=0,1,\dotsc,N-1$ are eigenvector of $\phi(Z)$ with eigenvalue $c\omega^k$. These eigenvalues are distinct roots of 1 and hence $S$ is linearly independent and a basis of $V$. We can reason similarly for $\phi(X)$. The converse is trivial. If $\gamma^k=I$ for $k<N$ then $\phi$ cannot be faithful. 

Next we recall some facts from the theory of characters associated with representation of a group \cite{Serre}. If $\rho$ is a representations of a finite group $G$ on a finite-dimensional vector space $V$, the character $\chi_\rho$ is a scalar function on $G$ defined by $\chi_\rho(g)=\tr(\rho(g))$. It is constant on conjugacy classes. If we have two characters $\chi_\rho$ and $\chi_{\rho'}$ corresponding to representations $\rho$ and $\rho'$ then their scalar product is defined as $(\chi_\rho,\chi_{\rho'})= (1/N)\sum_{g\in G}\conj{\chi_\rho(g)}\chi_{\rho'}(g)$. It is a fundamental result that two {\em irreducible} representations $\rho$ and $\rho'$ are (unitarily) equivalent if and only if $(\chi_\rho,\chi_{\rho'})\neq 0$. We apply this to the representation $\phi$ and $\phi'$ of $\heis{N}$. First, observe that since 
\(ZX^mZ^nZ^{-1}=\gamma^m X^mZ^n\), $\chi_\phi(X^mZ^n)=\omega^m\chi_\phi (X^mZ^n)$ which is possible iff either $m=0$ or $\chi_\phi(X^mZ^n)=0$. Conjugating with $X$ we conclude that $\chi_\phi$ is nonzero only on the centre of $\heis{N}$. Hence, to prove equivalence of $\phi$ and $\phi'$ it suffices to show the scalar product of $\chi_\phi$ and $\chi_{\phi'}$ is non-zero. But $\phi$ and $\phi'$ have the same effect on the center (generated by $\gamma$) of $\heis{N}$. Hence $(\chi_\rho,\chi_{\rho'})=(1/N)\sum_k  \conj{\chi_\phi(\gamma^k)}\chi_{\phi'}(\gamma^k)=1$. Since $\phi$ and $\phi'$ are equivalent there exists a unitary map $\Omega:V\rightarrow V$ such that $\phi'(g)=\hconj{\Omega}\phi(g)\Omega$. In particular, $\phi'(Z)=\phi(\tau Z)=\phi(X)=\hconj{\Omega}\phi(Z)\Omega$. Now let $\{\ket{j}:j=0,\dotsc, \}$ be a complete set of eigenvectors of $\phi(Z)$ with $\phi(Z)\ket{j}=\omega^j\ket{j}$ and similarly let $\{\ket{\tilde{j}}\}$ be an eigenbasis of $\phi(X)$. Then $\phi(Z)=\sum_j\omega^j\pj{j}$ and $\phi(X)=\sum_j\omega^j\pj{\tilde{j}}$. Observing that $\{\prj{j}{\tilde{k}}:j,k =0,\dotsc,N-1\}$   form a basis the space of operators on $V$ it is easy to check that $\Omega=\sum_j\prj{j}{\tilde{j}}$. We have also seen that $\phi(X)\ket{j}=\ket{j+1}$. From these and the normalization $\inp{\tilde{j}}{0}=1$ we get $\inpr{j}{\Omega}{k}=\omega^{-jk}/\sqrt{n}$. We have proved item 2. 

To prove the last assertion we again start with an eigenvector $\alpha$ with eigenvalue $a$ of $\psi({\bf z})$. Note that we can no longer assume that $a$ is an $K$th root of 1. However, the hypotheses that order of $\psi(\gamma)$ is $K$ implies that $\alpha, \psi({\bf x})\alpha,\dotsc, \psi({\bf x}^{K-1})\alpha$ are eigenvectors of  $\psi({\bf z})$ with {\em distinct} eigenvalues $a\psi(\gamma)^j$, $j=1,\dotsc,K-1$. They must be then independent. This implies ${\bf x}^K\alpha=\alpha$ and hence ${\bf x}^K=1$. Hence the eigenvalues of ${\bf x}$ must be $K$th roots of 1. Interchanging the role of ${\bf x}$ and ${\bf z}$ we conclude that $a$ must be a primitive $K$th root of 1 and the assertion follows.
\end{proof}
Note that the condition on $\gamma$ (maximal order) is necessary in case of the group $\heis{}$ and $\heis{N}$. For example, let $N=3$, $\rho(\gamma)=-I$, $\rho(Z)\ket{0}=\ket{0},\rho(Z)\ket{1}=-\ket{1}$, $\rho(Z)\ket{2}=\ket{2}$ and $\rho(X)$ the cyclic permutation. Then $\rho$ is an irreducible representation of $\heis{9}$. We see the connection between representations of the Heisenberg groups and the QFT. For a vector $\alpha$ we write $\tilde{\alpha}=\Omega\alpha$ for its Fourier transform. The Plancherel formula \(\norm{\alpha}^2=\norm{\tilde{\alpha}}^2\) is simply stating that the Fourier operator $\Omega$ is unitary. We also note that since the representations of the group $\heis{}$ is equivalent to $\heis{N}$ when $\gamma$ acts maximally it will be sufficient to consider $\heis{N}$ in a fixed representation space. However, when we are dealing with different representations (for example, taking tensor products) we have to deal with the full Heisenberg group. The condition that $o(\gamma)=\dim(V)$ is special case of general irreducible representations of $\heis{}$. It is sufficient for our purposes and will be implicitly assumed. Henceforth, for a fixed representation $\rho$ we simply write the action of a group element $g$ as $g\alpha$ instead of $\rho(g)\alpha$ if the context is clear. 

Now we turn our attention to distribution functions. We have seen that the distribution functions can be given an alternative characterization in terms of phase-point operators. The formula \eqref{eq:phaseOp} for these operators implies that they are linear combination of the unitary operators of the group. Thus, we look for them in the {\em group algebra}. Recall that for a group $G$ the group algebra $C(G)$ over complex numbers is the set of formal finite linear combinations $\sum_i c_ig_i,\; g_i\in G$ and $c_i\in \comp$. The algebra product is defined as 
\[ 
\sum_i c_ig_i\sum_j d_j g_j= \sum_{ij} c_id_jg_ig_j
\]
Any representation of the group is a representation of the group algebra and vice versa. Now for a {\em unitary}  representation $\rho$ of $G$ on a finite-dimensional vector space the character $\chi_\rho$ of the representation induces a scalar product on $C(G)$. Thus 
\beq\label{eq:involution}
(\mu, \nu)= \chi_\rho(\mu^*\nu) \text{ where } \mu =\sum c_ig_i,\nu=\sum c_j'g_j' \text{ and }\mu^*= \sum\conj{c}_ig_i^{-1}
\eeq
\def\bx{\bf x}
\def\bz{\bf z}
This is indeed a scalar product on $C(G)$. The resulting norm coincides with the Hilbert-Schmidt norm on the corresponding operators on $V$. Call an element $\mu\in C(G)$ self-adjoint if $\mu^*=\mu$. Let $G=\heis{N}$ or $\heis{}$. Since the central element $\gamma$ acts as a scalar we write elements of $C(G)$ in the form \(\sum_{m,n}c_{mn}X^mZ^n\). For a representation $\phi$ of $\heis{N}$ with $\gamma$ acting as $\omega I$ consider the following elements 
\[
A(x,z)=\sum_{mn} c_{mn}\omega^{-mx+nz}X^mZ^n,\;x,z=0,\dotsc,N-1\text{ in }C(\heis{N})
\]
We demand that the set $\cali{G}=\{A(x,z)\}$ be mutually orthogonal, self-adjoint and satisfy the following: the elements $P(x)=\sum_z A(x,z)$ and $Q(z)=\sum_x W(x,z)$ are projections, that is, $P(x)^2=P(x)$ and $P(z)^2=P(z)$. We call such a set of elements in $C(G)$ a Wigner set. 
We have the following theorem. 
\begin{thm}\label{thm:existWigner2}
For a representation $\phi$ of $\heis{N}$ on $N$-dimensional space $V$, Wigner sets exist in $C(\heis{N})$. If $\cali{G}$ is a Wigner set then for $A(x,z)\in \cali{G}$, $A(x,z)/N$ are phase-point operators. In other words, 
given a quantum state $\rho$ the function $W(x,z:\rho)\equiv \tr(\phi(A(x,z))\rho)/N$ is a distribution function. Conversely given a distribution function $W(x,z:\rho)$ on $V$ there is a unique Wigner set $A(x,z)$ in $C(\heis{N})$ such that $W(x,z:\rho)\equiv \tr(\phi(A(x,z))\rho)/N$. Wigner sets are translation invariant in the sense that the transformation $c_{mn}\rightarrow c_{mn}\omega^{am+bn}$, $a,b\in Z_N$ permutes the operators $A(x,z)$ in a Wigner set. 
\end{thm} 
\begin{proof}
The proof is similar to that of Theorem \ref{thm:existWigner1}. We only sketch some of the basic arguments since we are dealing with group algebras. First, the self-adjoint property implies conditions like \eqref{eq:R1} with $c_{mn}$ in place of $f(m,n)$ since the $\{X^mZ^n\}$ are independent in $C(\heis{N})$. Let us compute the scalar product of two elements from $\cali{G}$. Assuming now self-adjointness we have 
\[ 
\begin{split}
&(A(x',z'),A(x,z))=\\ &\tr\left(\sum_{\substack{m,n\\m',n'}}\conj{c}_{m'n'}c_{mn}\omega^{mn'}X^{m+m'}Z^{n+n'}\omega^{-(m'x'+n'z')}\omega^{-(mx+nz)}\right)\\
&=\tr\left(\sum_{\substack{m,n\\m',n'}} c_{mn}c_{N-m,N-n}\omega^{mn}\omega^{-m(x-x')-n(z-z')}\right)\\
&= N\sum|c_{mn}|^2\omega^{-m(x-x')-n(z-z')}\\
\end{split}
\]
In deriving the second step we use the fact that $\tr(\phi(X^jZ^k))=0$ unless $j=k=0 \mod N$. The last expression will be proportional to $\delta_{xx'}\delta_{zz'}$ if $|c_{mn}|^2=K$, a constant. We will fix $K$ shortly. Hence, we assume that $c_{mn}=K\omega^{b_{mn}}$. For the last requirement we  have 
\[
Q(z)=\sum_x A(x,z)= N\sum_n c_{0n}Z^n\omega^{-nz} = 
Q(z)^2= N^2\sum_{mn} c_{0m}c_{0n} Z^{m+n}\omega^{-z(m+n)}
\]
This would be possible if $c_{0n}=K=1/N$ for all $n$. We have already proved the existence of functions satisfying these conditions in Theorem \ref{thm:existWigner1}. The fact that $W(x,z:\rho)=\tr(A(x,z)\rho)$ is real follows from self-adjointness. The orthogonality property implies R2 in Section \ref{sec:propDist}: 
\[\tr(\rho\rho')=\sum_{xz} W(x,z:\rho)W(x,z:\rho').  \]
Finally, the property about marginals is equivalent to showing that $P(z)$ and $P(x)$ represent projections on $\ket{z}$ and $\ket{\tilde{x}}$ respectively. We can deduce this directly from the fact that 
\[ 
\phi(Z)=\sum_{j}\omega^j\pj{j} \text{ and } \phi(X)=\sum_{j}\omega^j\pj{\tilde{j}}
\]
The proof of the converse is straightforward. 

To prove the last statement let $c'_{mn}=c_{mn}\omega^{am+bn}$. Then 
\[ 
\begin{split}
A'(x,z)&\equiv \sum_{mn} c'_{mn}\omega^{-mx+nz}X^mZ^n\\
&= \sum_{mn} c_{mn}\omega^{-m(x-a)+n(z-b)}X^mZ^n\\
&= A(x',z') \text{ where } x'=x-a\text{ and } z'=z-b \\
\end{split}
\]
The assertion follows from this and the proof is complete. 
\end{proof}
We see the correspondence between orthogonal sets in the group algebra $C(\heis{N})$ and distribution functions. We have seen that the QFT arises out of a particular automorphism $\tau$ of the Heisenberg group. So we expect the general automorphisms of $\heis{N}$ and $\heis{}$ to contain more structure and information relating to QFT and distribution functions. It is easy to see that any two representations of $\heis{N}$ in which $\gamma$ acts maximally and has the same value are equivalent. In particular, if $\sigma$ is an automorphism of $\heis{N}$ that fixes $\gamma$ and $\phi$ is an arbitrary representation then $\phi$ and $\phi\cdot\sigma$ are equivalent. If $\{A(x,z): 0\leq x,z\leq N-1\}$ is a Wigner set then $\{\sigma A(x,z)\}$ is also a Wigner set. So we can generate new Wigner sets by automorphisms. As the value of $\sigma$ on $X$ and $Z$ determines it on $C(\heis{N})$ let $\sigma(X)=X^aZ^b$ and $\sigma(Z)=X^cZ^d$. We must have 
\[ 
\begin{split}
&(X^aZ^b)^N=\gamma^{abN(N-1)/2}=(X^cZ^d)^N=\gamma^{cdN(N-1)/2}=e\text{ and }\\
&X^cZ^dX^aZ^b=\gamma^{ad-bc}X^aZ^bX^cZ^d=\gamma X^aZ^bX^cZ^d
\end{split}
\]
for $\sigma$ to be an automorphism. The second condition implies that the matrix 
\[
M^\sigma\equiv  \begin{pmatrix}
a & b \\
c & d \\
\end{pmatrix}
\]
has determinant 1. That is, $M^\sigma\in SL(2, Z_N)$, the set of matrices with entries in the ring $Z_N$ and determinant $1\mod{N}$. Conversely, given $M\in SL(2,Z_N)$, $N$ odd we can define an automorphism $\sigma_M$ as above. If $N$ is even this simple definition of $\sigma_M$ does not work in general. For example, if $ab$ is {\em odd} then $(X^aZ^b)^N=\gamma^{N/2}=-1$. This is reminiscent of half-integral representation of rotation group ($SU(2)$ actually). Hence for even $N$ we have an automorphism of $\heis{}$ rather than $\heis{N}$. Note that 
in this case for any $M\in SL(2,Z_N)$, $(\sigma_M(g))^{2N}=1,\; g\in \heis{N}$. There is however a proper subgroup of $SL(2,Z_N)$ which induces an automorphism of $\heis{N}$. Alternatively, for even $N$ define the function 
\[\sgn(u)=\begin{cases} 0,\;  u\in Z \text{ and } u \text{ even }\\
1,\; u \text{ odd }
\end{cases}
\]
Now we can define the automorphism $\sigma_M$ on $\heis{N}$, $M\in SL(2,Z_N)$ for odd $N$ and for even $N$ we define it on the representation space. 
\beq \label{eq:induced_auto}
\sigma_M(X)=\begin{cases} X^aZ^b, \; N \text{ odd }\\
\omega^{\sgn(ab)/2}X^aZ^b, \; N \text{ even }
\end{cases}
\eeq
Let $A(x,z)$ be a Wigner set and $\phi$ a representation of $\heis{N}$. Then we have seen that a distribution function is defined by 
\[ W(x,z:\rho)= \tr(\phi(A(x,z))\rho)/N\]
This means that if $\phi$ and $\phi'$ are equivalent representations connected by a a unitary operator $U$ and $W$ and $W'$ are the corresponding distribution functions then 
\beq\label{eq:stateTransform}
W(x,z:\rho)= W'(x,z:\hconj{U}\rho U)
\eeq
In particular, if $M\in SL(2,Z_N)$ then it induces an equivalent representation $\phi_M$. In case $N$ is odd $\phi_M$ is the representation that is given via the automorphism generated by $M$. In even dimension $\phi_M$ is defined by \eqref{eq:induced_auto} above. Now let us look at other ``marginals'' of a distribution function $W(x,z:\rho)$. One way of constructing such marginals is via a finite Radon transform \cite{Diaconis,Fill}. Thus for $f:Z_N\times Z_N\rightarrow \comp$ define ``lines'' 
\beq\label{eq:radon}
\begin{split}
&S_{ab}=\bigl\{(x,z)\in Z_N\times Z_N: ax+bz= 0\bmod{N},\; \gcd{(a,b,N)}=1\bigr\}\\
& \widehat{W}(x',z':\rho) = \sum_{x,z\in S_{ab}+(x',z')}W(x,z,:\rho) 
\end{split}
\eeq
The condition $\gcd{(a,b,N)}=1$ ensures that the ``line'' $ax+bz=t$ has a solution for all $t\in Z_N$. The ``coordinate axes'', for example, correspond to the sets $S_{10}$ and $S_{01}$. The function $\widehat{W}(z')$ in \eqref{eq:radon} is a {\em Radon transform} \cite{Fill} of the distribution function $W(x,z)$ and each pair $(a,b)\in Z_N\times Z_N$ such that $\gcd{(a,b)}$ is invertible in $Z_N$,  defines such a transform. Let $x'=ax+bz$. Let $(c,d)\in Z_N\times Z_N$ such that the matrix
\[
M=\begin{pmatrix} a & b\\ c & d \end{pmatrix} \in SL(2,Z_N)
\]
and set $z'=cx+dz$. In the following it will be convenient to use vector notation. Thus  $\pmb{\xi}=(x,z)^T\in Z_N\times Z_N$ is a two-dimensional ``vector'' \footnote{To be accurate $Z_N\times Z_N$ is a {\em module} over the ring $Z_N$.}. We will also occasionally use the component notation: $\pmb{\xi}= (\xi_1, \xi_2)^T$. So the distribution functions and phase-point operators may be written as $W(\pmb{\xi}:\rho)$ and $A(\pmb{\xi})$ respectively. Then a marginal with respect to the second component is given by 
\[
\widehat{W}(z':\rho)= \sum_{M\pmb{\xi}_1\in Z_N} W(M\pmb{\xi}:\rho)
\]
Here $z'=M\pmb{\xi}_2$. In analogy with the continuous case we require that $\widehat{W}(z':\rho)$ be a probability distribution with respect to $z'$. More precisely, in the representation $\phi_{M^{-1}}$ of the Heisenberg group corresponding to the automorphism induced by $M^{-1}$,  $\widehat{W}(z';\rho)$ gives probability distribution of the {\em quantum observable} $-i\ln\phi_{M^{-1}}(Z)$ in the state $\rho$. However, there is a sharp difference between the distribution functions in even and odd dimensions. The general marginal condition holds in odd dimensions for the Wigner distribution function defined by \eqref{eq:order1} but not for even dimensions. For even dimension we have more complicated formulas for the marginals of the Wigner function. In fact we will show that in this case no distribution function satisfying conditions R1-R4 in subsection \ref{sec:propDist} will satisfy the general marginal condition for all $M\in SL(2,Z_N)$. 
\begin{thm}\label{thm:Radon1}
Let $\phi$ be a representation of $\heis{N}$ on $V$. Let   
\beq
A(x,z)= \frac{1}{N^2}\sum_{m,n} f(m,n) X^mZ^n \omega^{-mx+nz} 
\eeq
be a Wigner set and $W(x,z:\rho)=\tr(A(x,z)\rho)$ be the corresponding distribution function. 
\[
M=\begin{pmatrix} a & b\\ c & d \end{pmatrix} \in SL(2,Z_N)
\]
We say that simple marginal condition (with respect to $M$) is satisfied if the marginals 
\beq\label{eq:condMarginal1} 
\widehat{W}(z':\rho)\equiv \sum_{M^{-1}\xi_1\in Z_N} W(M^{-1}\pmb{\xi}:\rho)
\eeq
are probability distribution in the eigenbasis of the operator $\phi_M(Z)\equiv \phi(\sigma_M(Z))=\phi(X^cZ^d)$. Then the following statements hold. 
\be
\item
The (simple) marginal condition is satisfied if and only if 
\[ A_M(x,z)\equiv A(M^{-1}\pmb{\xi})\]
is a Wigner set for the representation $\phi_M$. 
\item
If marginal conditions are satisfied for all $M\in SL(2,Z_N)$ then the Wigner sets (and distribution functions) are determined uniquely up to translations. 
\item
If the dimension $N>2$ is even it is not possible to satisfy the marginal condition for all $M\in SL(2,Z_N)$. 
\item
In odd dimension Wigner functions are given by 
\beq\label{eq:WignerBasicOdd}
 W(x,z)= \sum_{m,n} \omega^{mn(N+1)/2}X^mZ^n \omega^{-(mx+nz)}
\eeq 
up to translations. 
In that case $\widehat{W}(z':\rho)=\inpra{\alpha_{z'+cd/2}}{\rho}$ where $\ket{\alpha_j}$ are the eigenvectors of $\sigma_M(Z)$. 
\ee
\end{thm}
\begin{proof}
Since $\gcd{(a,b,N)}=1$ there exist integers $c,d\text{ and }k$ such that $ad-bc+kN=1$  and hence $ad-bc= 1 \bmod{N}$. For a matrix $M$ let $M'=(M^T)^{-1}$. The first assertion in the list is relatively straightforward. If 
the equation \eqref{eq:condMarginal1} is satisfied then $W_M(x,z)=W(M^{-1}x,M^{-1}z)$ is a distribution function. The condition R1 (reality) is clear, R2 follows from the state transformation equation \eqref{eq:stateTransform} and \eqref{eq:condMarginal1} gives the marginal condition R4. From the correspondence between distribution functions and Wigner sets the first assertion is clear. 

Suppose the marginal conditions are satisfied for some $M\in SL(2, Z_N)$ given above. Using the formula
\beq\label{eq:useful}
(X^uZ^v)^m= \omega^{uvm(m-1)/2}X^{um}Z^{vm}
\eeq
we get setting $\pmb{\zeta}= (m,n)^T\in Z_N\times Z_N$ 
\[
\begin{split}
&N^2\sum_{\xi_1} A(M^{-1}\pmb{\xi}:\rho)= \sum_{M\xi_1}\sum_{m,n} f(m,n)X^mZ^n\omega^{-\pmb{\zeta}\cdot M^{-1}\pmb{\xi}}\\
&=\sum_{\xi_1}\sum_{m,n} f(m,n)X^mZ^n\omega^{-M'\pmb{\zeta}\cdot \pmb{\xi}}= \sum_{M\xi_1}\sum_{\zeta'=M'\zeta} f(M^T\pmb{\zeta}')X^mZ^n\omega^{-\pmb{\zeta}'\cdot \pmb{\xi}}\\
&= \sum_{x\in Z_N}\sum_{m,n} f(am+cn,bm+dn)\omega^{-(abm(m-1)/2+cdn(n-1)/2+bcmn)}\times\\
&\quad\quad\quad\quad\quad{\sigma_M(X)^{m}\sigma_M(Z)^{n}}\omega^{-(mx+nz)},\; (\xi_1=x)\\
&= \sum_n f(cn,dn)\omega^{-(cdn(n-1)/2)}\sigma_M(Z)^{n}\omega^{-nz}\\
&\equiv \sum_n g(n)Z'^n,\; g(n) = f(cn,dn)\omega^{-(cdn(n-1)/2)}\text{ and } Z'= \sigma_M(Z)\omega^{-z}\\
\end{split}
\]
We require that the operator $T=\sum_n g(n)Z'^n$ be a projection: it must be hermitian and satisfy $T^2=T$. Hence, we must have 
\[
\sum_{m,n} {g(m)} g(n) {Z'}^{n+m}= \sum_k\sum_m {g(m)}g(k-m)Z'^k=\sum_k g(k)Z'^k
\]
Since $Z'$ like $Z$ has no repeated eigenvalue its minimum polynomial is the characteristic polynomial $\lambda^N-1$. Hence the operators $I,Z',Z'^2,\dotsc, Z'^{N-1}$ are linearly independent and we must have 
\[ \sum_m {g(m)}g(k-m)= g(k)\]
But the left side is the {\em convolution} of the function $g$ with itself. Taking (finite) Fourier transform of both sides we get $\tilde{g}^2=\tilde{g}$. Then $\tilde{g}(m)=1\text{ or } 0$. This implies that 
\[ g(n)= \sum_i \omega^{t_in}\]
where $t_i\in Z_N$ are the values at which $\tilde{g}=1$. But from condition R2 we infer that $|g(x)|=1$ and since R1 implies that $g(0)=1$ we conclude that there must be exactly one term in the above sum: 

\beq\label{eq:basicRecur}
 g(n)= f(cn,dn)\omega^{-cdn(n-1)/2}=\omega^{t(c,d)n}
\eeq
Putting $n=1$ this implies that $f(c,d)=\omega^{t(c,d)}$ whenever $\gcd(c,d,N)=1$. Hence we rewrite the above equation as 
\beq\label{eq:basicRecur1}
 f(cn,dn)= f(c,d)^n\omega^{cdn(n-1)/2} \text{ and so } f(cn,n)= f(c,1)^n\omega^{cn(n-1)/2} 
\eeq
Now suppose $N$ is even. By the definition of Wigner sets they must be independent since the operators are mutually orthogonal. Consequently, the function $f$ must be periodic with period $N$ and since $f(c,1)=\omega^{t(c,1)}$, $t(c,1)$ must be an integer. Putting $n=N$ in the second equation in \eqref{eq:basicRecur1} and noting that $f(x,0)=1,\;\forall x\in \mathbb{Z}$ we get a contradiction when $c$ is odd for the right side is $-1$. Hence it is not possible to have Wigner sets satisfying {\em all} simple marginal conditions. 

Next suppose that $N$ is odd. Then 2 has an inverse $(N+1)/2$ in $Z_N$. It is an easy verification that the function 
$f(m,n)=\omega^{mn(N+1)/2}$ satisfies the functional relation \eqref{eq:R1}. To prove uniqueness we assume that $t(m,n)$ can be extended to all $Z$ and that it can be expressed as a polynomial in $m$ and $n$ with integer coefficients (which may depend upon $N$). Since $f(m,0)=f(n,0)=1$ we may assume that the polynomial is of the form $t(m,n)=\omega^{mn[a_0+g(m,n)]}$ where $a_0$ is a constant and $g(m,n)$ is a polynomial without constant term. Then we have 
\[ f(cn,n)=\omega^{cn^2(a_0+g(cn,n))}= \omega^{nc(a_0+g(c,1))}\omega^{cn(n-1)/2}\]
Since this must be satisfied for all $n$ we must have $a_0=(N+1)/2$ and $g=0\mod{N}$. This proves uniqueness up to linear terms. 

The last statement is easily derived from the above proof of existence and uniqueness of distribution function satisfying all the marginal conditions for odd $N$. 
\end{proof}

We note that a similar relation holds for the marginal distribution over $x$ when we average over the variable $z$. In fact, satisfaction of marginal conditions under the full $SL(2,Z_N)$ for one variable implies the same for other. 
In even dimensions there exist no distribution function satisfying all marginal conditions. Therefore, we have to relax some of the conditions of the theorem to get the marginal distributions. Let us recall why the marginal conditions are desirable. One of the main reasons is that by determining sufficient number of marginal distributions we can reconstruct the state. The simple marginal condition stated in the Theorem \ref{thm:Radon1} is satisfied (see \eqref{eq:condMarginal1} and the statement that follows it) the marginal distribution corresponds to probabilities for a complete projective measurement in a suitable basis. In even dimension we have three options. 

\be
\item
We do not require that the Wigner set be {\em independent}. Then the representation of the Heisenberg group $\heis{N}$ need not be irreducible. This was the approach adopted in \cite{Leonhardt96}. 
\item
We drop the conditions that the marginals be simple type. As will be shown next we can still determine the ``marginal'' distributions from the measurement probabilities. 
\item
We do not demand that the marginal condition be satisfied for the full $SL(2,Z_N)$ but only for a subset. We show that in case $N=2^K$ there is such a subset and the marginal distribution for it are sufficient to reconstruct the distribution function. 
 \ee
We start with the first option \cite{Leonhardt96,Miquel}. Since the operators $A(x,z)$ are no longer independent the function $f$ (as a function on $\mathbb{Z}$) is not required to be periodic and the labels $(x,z)$ can take any integer values. A minimal extension is obtained by looking at the basic recurrence relations \eqref{eq:basicRecur1}. The problematic factor $\omega^{cn(n-1)/2}$ is periodic with a period $2N$ (as function on $\mathbb{Z}$). The same relations then suggest that  we take $f(m,n)=\omega^{mn}/2$ where $\omega^{1/2}$ is a primitive $2N$th root of 1. Hermiticity of phase-point operators then require that we now define them as 
\[ A(x,z) = \sum_{m,n\in Z_{2N}} \omega^{mn/2}X^mZ^n\omega^{-(mx+nz)/2}\]
Because of redundancy these operators are not uniquely determined (up to linear factors). But we can modify the proof in Theorem \ref{thm:Radon1} for odd dimension to determine the possible solutions in this case. 

Next we look at option 2. We defined a family of distribution functions say $W(x,z:\rho,\nu)$ in the even case in \eqref{eq:order1} depending on some function $\nu$. The function $\nu$ is arbitrary apart from the condition \eqref{eq:extraPhase_even}. Let $W_0$ denote the special case when $\nu$ is given by \eqref{eq:evenSpecial}. Of course, $W_0$ does not satisfy the marginal condition but the results below show how it may be computed from the measurement probabilities. 
\begin{propn}\label{prop:Radon2}
Let $V$ be an irreducible representation space of $\heis{N}$, $N$ even. Let 
\[ M= \begin{pmatrix} a & b\\ c & d \end{pmatrix}\in SL(2,Z_N)\]
Let $u=\gcd(t,N)$ where $t=c$ if $c$ even and $d$ otherwise. Suppose $N/u$ is even. Then 
\beq \label{eq:Wigner_marginalEven1}
\begin{split}
\widehat{W}(z':\rho,\nu)&\equiv \sum_{x'\in Z_n} W(M^{-1}x', M^{-1}z':\rho)\\
&=\sum_j\rho_{jj} \sum_n \nu_{cn,dn} (-1)^{dn\bfloor{\frac{cn}{N}}+cn\bfloor{\frac{dn}{N}}}\omega^{((cd+\sgn(cd))/2+j-z')n}
\end{split}
\eeq
where $\bfloor{x}$ is the greatest integer $\leq x$. In particular, for $W_0$ with $\nu_{mn}$ given by \eqref{eq:evenSpecial} we have 
\beq\label{eq:Wigner_marginalEven2}
\begin{split}
& \widehat{W_0}(z':\rho)= \\
&\bigl(\inpra{\alpha_{z'-\frac{cd-\sgn(cd)}{2}}}{\rho}+\inpra{\alpha_{\frac{N}{2}+z'-\frac{cd-\sgn(cd)}{2}}}{\rho}\bigr)/2\\
&-\frac{2}{N} \sum_j\rho_{jj}\sum_{n\text{ odd}}(-1)^{d\bfloor{\frac{cn}{N}}+c\bfloor{\frac{dn}{N}}}h(j,z')\\
&\text{ where } h(j,z')=\begin{cases} \cos\frac{2\pi(\frac{cd-1}{2}+j-z')}{N} \text{ if } cd \text{ odd } \\
 -\sin\frac{2\pi(\frac{cd-1}{2}+j-z')}{N} \text{ if } cd \text{ even } 
 \end{cases}
 \\
\end{split}
\eeq
\end{propn}

The proof is given in the appendix. Observe that the if $N=2^k,\;K>1$ then $N_1=N/u$ is always even. We can also write the appropriate formulas for the case $N_1$ odd. We avoid doing so as they are even more complicated. We can also simplify the trigonometric sums in \eqref{eq:Wigner_marginalEven1}. However, note that if we know the probabilities $\inpra{j}{\rho}$ then in principle the Radon transform $\hat{W}(z:\rho,M)$ can be computed by evaluating these sums. In case of odd dimension the expressions for the marginals are simpler but we still have to estimate the probability distribution in the basis $\{\ket{\alpha_j}\}$ defined above. If we have these the probabilities doing the sums in the even case is routine. So, is there a deeper reason for imposing the marginal conditions on the distribution function? Two possible reasons could be simplicity and some theoretical insight. 

We consider the second option listed above for dimension $N=2^k$ only. Thus we aim to construct a distribution function which satisfies the marginal conditions for only a subset of $SL(2,Z_N)$. 
The theorem below gives an explicit formula for this important case. Thus let $L_1\subset SL(2,Z_N)$ be a subset consisting of the following matrices. If $M\in L_1$ then each row has at least one entry $= 1$ and if the diagonal entry is $\neq 1$ it is even. 
\begin{thm}\label{thm:WignerQubit}
Let $N=2^k$. Define 
\beq\label{eq:WignerFuncQubit}
\begin{split}
W_1(x,z)= &\frac{1}{N^2}\Bigl(\sum_{\substack{m,n\\\text{ even}}}\ave{X^mZ^n} \omega^{mn/2-(mx+nz)}+ \sum_{m }\ave{X^m} \omega^{-mx}+\sum_{n }\ave{Z^n} \omega^{-nz}+\\
&\sum_{\substack{m>0\\n \text{ odd }}} (-1)^{\bfloor{\frac{(mn^{-1})n}{N}}}\ave{X^mZ^n} \omega^{mn/2-(mx+nz)}
+\\
&\sum_{\substack{n\text{ even }>0\\m \text{ odd }}} (-1)^{\bfloor{\frac{(nm^{-1})m}{N}}}\ave{X^mZ^n}  \omega^{mn/2-(mx+nz)}\Bigr)\\
\end{split}
\eeq
where the expressions like $(mn^{-1})$ are first computed modulo $N$ in the residue class $\{0, \dotsc, N-1\}$ and then treated as an integer. $\floor{x}$ denotes the largest integer less than or equal to $x$. 
Then $W_1$ satisfies the conditions {\em R1-R4} and for every $M\in L_1$, $W_1$ satisfies a simple marginal condition with respect to the variable $x$: 
\beq\label{eq:Wigner_marginalQubit}
\begin{split}
\widehat{W_1}(z':\rho,\nu)& \equiv \sum_{x'\in Z_n} W_1(M^{-1}x', M^{-1}z':\rho)\\ 
&=\begin{cases} \inprab{\alpha_{z'-\frac{c+\sgn(c)}{2}}}{\rho},\; d=1\\ 
\inprab{\alpha_{z'-\frac{d}{2}}}{\rho},\; c=1\text{ and } d\text{ even}
\end{cases}\\
\end{split}
\eeq 
\end{thm}
\begin{proof}
We first note that the notation $n^{-1}$ makes sense in the ring $Z_N$ since every odd $n$ is invertible. 
The reality condition R$_1$ is seen from the following simple observation. For $0\leq m,n <N$ let $mn^{-1}=k_1N+n_1$, $k_1\geq 0$ and $0\leq n_1<N$. Since $(N-m)(N-n)^{-1}=(mn^{-1})\bmod N$ we get 
\[((N-m)(N-n)^{-1})(N-n)= (m-1-k_1)N+n_1\]
This implies that $k_1$ has same (opposite) parity as $\bfloor{((N-m)(N-n)^{-1})(N-n)}$ if $m$ is odd (even). Hence
\[ (-1)^{\bfloor{(mn^{-1})n}}= (-1)^{m+n}(-1)^{\bfloor{((N-m)(N-n)^{-1})n}}\]
We can argue similarly for $m$ odd and $n$ even. Hence the reality condition \eqref{eq:R1} is satisfied. It is clear that $W_1$ is normalized. The other conditions easily follow from the definition and the analysis of these conditions in Section \ref{sec:propDist}. Finally, the simple marginal condition with respect to $x$ is seen to be  satisfied as follows. From Theorem \ref{thm:Radon1} and Proposition \ref{prop:Radon2} we note that we have to consider pairs of the form $(cn,dn)$, where $(c,d)$ is the second row of $M$, in the calculation of the marginals. Using the notation of Proposition \ref{prop:Radon2} we set 
\[ \nu_{mn}=\begin{cases} 1 \; m,n \text{ even }\\ 
\bfloor{\frac{(mn^{-1})n}{N}}\; n \text{ odd }\\
\bfloor{\frac{(nm^{-1})m}{N}}\; m \text{ odd },n \text{ even }
\end{cases}
\]
As the matrices belong to $L_1$ we consider two cases. If the diagonal element $d=1$ then the only terms in the sum yielding $W_1$ that contribute to the marginal are indexed by $((cn),n)$ where $n$ runs through $Z_N$ and $(cn)$ is calculated $\bmod N$. The case $c=0$ is already covered. If $c\neq 0$ then from \eqref{eq:Wigner_marginalEven1}
\[ 
\begin{split}
&\widehat{W_1}(z':\rho,\nu)\equiv \sum_{x'\in Z_n} W_1(M^{-1}x', M^{-1}z':\rho)\\
&=\sum_j\rho_{jj} \sum_n \nu_{cn,dn} (-1)^{dn\bfloor{\frac{cn}{N}}+cn\bfloor{\frac{dn}{N}}}\omega^{((cd+\sgn(cd))/2+j-z')n}\\
&= \sum_j\rho_{jj} \bigl(\sum_{n\text{ odd}} \nu_{cn,n}\bfloor{\frac{cn}{N}}\omega^{((c+\sgn(c))/2+j-z')n}+\sum_{n\text{ even}}\nu_{cn,n}\omega^{((c+\sgn(c))/2+j-z')n}\bigr)\\
&= \sum_j\rho_{jj}\sum_n\omega^{((c+\sgn(c))/2+j-z')n}=\inprab{z'-\frac{c+\sgn(c)}{2}}{\rho}\\
\end{split}
\]
For the case $c=1$ and $d$ even the terms in which $n$ is odd drop out from the sum for 
$\widehat{W_1}(z':\rho,\nu)$ and the proof is similar to the first case. 
\end{proof}
We note that the subset $L_1$ of matrices from $SL(2,Z_N)$ cannot be extended arbitrarily preserving the property of simple marginals. For example, if we admit matrices with $c=1$ and $d$ {\em odd} then we get a factor of $\sgn(\bfloor{\frac{(d^{-1})n}{N}})$ instead of $\sgn(\bfloor{\frac{dn}{N}})$. The two need not be equal. However, as we will see below the set $L_1$ is sufficient to determine $W_1$. 
\subsection{Inverse Radon transform and state determination}\label{sec:inverseRadon}
In the previous section we have seen that the finite Wigner distribution function enjoys a rich variety of marginal properties. We can use this to determine the former. This is equivalent to inverting a finite set of Radon transforms. From the distribution function we can determine the state. The invertibility of the Radon transforms also shows that the Wigner distribution function is unique up to a translation. In the rest of the section $W(x,z)$ will denote the Wigner distribution function. Replacing the matrix $M^{-1}$ by 
\[
M= \begin{pmatrix} a & b \\ c & d \\ \end{pmatrix}, \quad \det{M}= 1\bmod{N} 
\]
 in Theorem \ref{thm:Radon1} we rewrite the basic Radon property stated in \eqref{eq:condMarginal1} and the statement following. For example, for odd $N$ 
\beq \label{eq:basicRadon2}
\widehat{W}(z:\rho, M)\equiv \sum_{x\in Z_N} W(Mx, Mz:\rho)= \tr(\pj{\alpha_{z-ac}}\rho)
\eeq
The problem is to reconstruct $W(x,z)$ from $\widehat{W}(z:\rho, M)$. Call the later the Radon transform of $W$ with respect to the matrix $M$. The idea is that $\widehat{W}(z:\rho, M)$ is the probability distribution of the observable $-i\ln{(X^{-c}Z^a)}$ in the odd case. In case of even dimensions it can be computed from the  distributions. Assuming that these distributions can be approximately determined experimentally we can reconstruct $W$ and hence $\rho$. We have seen that in odd dimension $N$ there is a distribution function satisfying simple marginal conditions for every $M\in SL(2,Z_N)$ and in dimension $N=2^k$ we have only a subset of $SL(2,Z_N)$ with simple marginal conditions. We give explicit formulas for these two cases. First some notation. For a subset $S$ of some set let $\chi_S$ denote the indicator function: $\chi_S(x)=1$ if $x\in S$ and 0 otherwise. In the rest of the section we use the boldface vector notation to denote a member of $Z_N\times Z_N$ and other non-bold letters to denote ``scalars'' belonging to $Z_N$. For example, 
\[{\bm \mu}= \begin{pmatrix} \mu_1\\\mu_2\end{pmatrix},\quad \mu_1,\mu_2\in Z_N\]
Given $M\in SL(2,Z_N)$ let ${\bf C}_i(M),\;i=1,2$ denote the column vectors of $M$. Let 
\[
S_i(M)= \{ {\bf C}_i(M)x: x\in Z_N\}\subset Z_N\times Z_N,\; i=1,2
\]

\begin{thm}\label{thm:InverseRadon}
Any distribution function $W(x,z)$ can be uniquely determined from the (finite) set of Radon transforms $\widehat{W}(z:\rho,M)$ where 
\[ M=\begin{pmatrix} a & b\\ c & d \end{pmatrix} \in SL(2,Z_N)\]
In particular, for odd dimensions the Wigner function given in \eqref{eq:WignerBasicOdd} 
and any $M$ we have  
\beq\label{eq:InverseRadon_odd}
\widetilde{W}(ct,-at)= \frac{1}{N}\sum_z \inpra{\alpha_{z-ac/2}(M)}{\rho}\omega^{zt}
\eeq
and for $N=2^k$ and $M\in L_1$ 
\beq\label{eq:InverseRadon_qubit}
\widetilde{W}(ct,-at)=\begin{cases} \frac{1}{N}\sum_z  \inprab{\alpha_{z'-\frac{c+\sgn(c)}{2}}(M)}{\rho}\omega^{zt},\; a=1\\ 
\inprab{\alpha_{z'-\frac{a}{2}}(M)}{\rho}\omega^{zt},\; c=-1\text{ and } a\text{ even}
\end{cases}
\eeq
where $\widetilde{W}$ is the Fourier transform of $W$ in $Z_N\times Z_N$ and $\ket{\alpha_j(M)}$ are the eigenvectors of $\sigma_{M^{-1}}(Z)=X^{-c}Z^a$.\footnote{Recall that here  $M$ replaces $M^{-1}$ of theorem \ref{thm:Radon1}.\label{fn:chngNot}} In either of the cases the Wigner function $W$ or $W_1$ can be reconstructed from the marginal distributions. 
\end{thm}
\begin{proof}
Write the Radon transform \eqref{eq:basicRadon2} as 
\[
\widehat{W}(z:\rho, M) = \sum_{x\in Z_N} W(Mx, Mz:\rho)= \sum_{{\bf x}\in S'_1(M)} W({\bf C}_2(M)z-{\bf x}: \rho)\\
\]
where $S'_1=-S_1$ We can write this as a {\em convolution}. Thus 
\[
\begin{split}
\widehat{W}({\bf u}:\rho, M)&= \chi_{S_z}({\bf u})\sum_{{\bf x}}W({\bf u}-{\bf x}:\rho)\chi_{S'_1(M)}({\bf x})\\
&= \chi_{S_z}({\bf u})( W\star \chi_{S_1'(M)} ({\bf u})),\quad S_z=\{{\bf C}_2(M)z\}\\
\end{split}
\]
Now we take the finite Fourier transform of the above equation in the group $Z_N\times Z_N$ \cite{Terras}. Recall that the Fourier transform of a complex function $f({\bf u})$ on $Z_N\times Z_N$ by $\tilde{f}$ is a function on the dual group $(Z_N\times Z_N)^*$: 
\[ \tilde{f}(\bm{\mu})=\frac{1}{N}\sum_{u_1,u_2}\omega^{-(\mu_1u_1+\mu_2u_2)}f({\bf u})\]
Using the fact that $\widehat{W}=\widehat{W}\chi_{S_z}$ and that the Fourier transform of a convolution is a product and vice versa we have (suppressing $\rho$ and $M$)
\[ 
\begin{split}
&\widetilde{\widehat{W}}(\bm{\mu})=\sum_{u_1,u_2}\widehat{W}({\bf u})\omega^{-\bm{\mu}\cdot {\bf u}}\\
&=\widetilde{\chi}_{S_z}\star(\widetilde{W}\widetilde{\chi}_{S'_1(M)})(\bm{\mu})= \sum_{\nu} \widetilde{\chi}_{S_z}(\bm{\mu}-\bm{\nu})\widetilde{W}(\bm{\nu})\widetilde{\chi}_{S'_1(M)}({\bm \nu})\\
&=\sum_{\{\bm{\nu}:a\nu_1+c\nu_2=0\}}\omega^{-[(\mu_1-\nu_1)b+(\mu_2-\nu_2)d]z}\widetilde{W}(\bm{\nu})\\
&= \omega^{-(\mu_1b+\mu_2d)z}\sum_t \omega^{((ct)b-(at)d)z}\widetilde{W}(ct,-at)\\
&= \omega^{-(\mu_1b+\mu_2d)z}\sum_t \omega^{-tz}{F}(t)= \sqrt{N}\omega^{-(\mu_1b+\mu_2d)z}\widetilde{F}(z)\\
\end{split}
\]
where $F(t)= \widetilde{W}(ct,-at)$ and $\widetilde{F}$ is its Fourier transform in $Z_N$. In proving the above we use the following facts: $\widetilde{\chi}_{S'_1(M)}(\bm{\nu})\neq 0$ iff $a\nu_1+c\nu_2=0$ and the solution to the congruence equation $a\nu_1+ c\nu_2= 0\bmod N$ is given by the set $\{(ct,-at):t\in Z_N$. This follows from a similar result for linear Diophantine equations \cite{Mordell69} and the  fact that $\gcd(a,c,N)=1$. We also use  $\det{M}=ad-bc=1$ in the las but one step. The factor $\sqrt{N}$ appears because of the normalization used in our definition of FFT. It now follows that 
\beq\label{eq:basicInverseRadon}
\widetilde{W}(ct,-at)= \frac{1}{N} \sum_z \widehat{W}(z)\omega^{zt}
\eeq
This formula is valid for {\em any} distribution function. Let now $N$ be odd. Combining this with the equation \eqref{eq:condMarginal1} in Theorem \ref{thm:Radon1} we get \eqref{eq:InverseRadon_odd}. 
 Similarly, when $N=2^K$ and $M^{-1}\in L_1$ we obtain \eqref{eq:InverseRadon_qubit}. Note that the formulas in 
 \eqref{eq:Wigner_marginalQubit} are valid under the assumption that $M\in L_1$ (see the footnote \ref{fn:chngNot} above).
 
We next show that it is always possible to find $a,c\in Z_N$ such that $\gcd{(a,c,N)}=1$ and the ``lines'' $\{(ct,-at):t\in Z_N\}$ cover the ``plane'' $Z_N\times Z_N$ in the two cases above. When $N$ is odd this is obvious. If $N=2^k$ consider $(x,y)\in Z_N\times Z_N$. For $0< j<N$ let $h_j$ denote the highest power of 2 that divides $j$, that is, $j/2^{h_j}$ is an odd integer. If $h_x\geq h_y$ then we put $a=1$ and $c=-2^{h_x-h_y}(y/2^{h_y})^{-1}$ where the inverse is evaluated in $Z_N$ and we assume that $y\neq 0$. Then $(x,y)=(ct,-at)$ for $t=-y$. If $h_x<h_y$ then put $c=1$ and $a=-2^{h_y-h_x}(x/h_x)^{-1}$. We have therefore shown that in all these cases the Radon transforms together can be inverted for from the values  $\widehat{W}(\mu_1,\mu_2)$ so obtained we can take the inverse Fourier transform and the last assertion of the theorem is proved. 
\end{proof}
We can thus recover any distribution function $W(x,z:\rho)$ and consequently the state $\rho$ from the Radon transform data which are in turn probability distribution of measurement in appropriate bases (see \eqref{eq:basicRadon2}). The theorem shows the existence of an inverse transform corresponding to the set of Radon transforms of $W$, each corresponding to an element $M$ in the group $SL(2,Z_N)$. But we do not need all the Radon transforms. What is an optimal subset $Q\subset SL(2,Z_N)$ that suffices to determine the state uniquely from probability distributions corresponding to measurements in appropriate bases? This question can only be satisfactorily answered in the context of prior information about the state. One can show that without any such information the cardinality of $Q$ is $O(N)$. Even then we have a lot of freedom. We can use our choices so as to ensure optimal measurement. Recall from Theorem \ref{thm:Radon1} that the Radon transforms are given by probability distribution (corresponding to a state $\rho$) in the basis that diagonalizes the unitary operator $X^cZ^d$. The only condition imposed on the pair $(c,d)\in Z_N\times Z_N$ is that $\gcd{(c,d,N)}=1$. We can often compute this basis explicitly. Then we can use quantum circuits to transform our original ``computational basis'' to the required basis. A criterion for the choice of $(c,d)$ could be those that minimizes the size of the circuit. For example, if $N=6$, the choice $c=3,d=2$ leads to a particularly simple basis. The analysis becomes simpler if the dimension $N$ is a prime power. We aim to address these issues in future. 
\subsection{Distribution functions and quantum information}
In this section we discuss some potential applications of distribution functions in quantum information processing (QIP). This is a developing area and we only sketch how our formalism may prove useful in various areas in QIP. For this it is best to view the distribution function as coefficients in the expansion of the state in some orthonormal basis in the space of operators, in particular, the basis consisting of phase-point operators. First we generalize to automorphism groups of the {\em group algebra} $C(\heis{N})$: a linear isomorphism $T:C(\heis{N})\rightarrow C(\heis{N})$ such that $T(xy)=T(x)T(y)$ is bijective. It is sufficient to check the last condition for the generators $X$, $Z$ and $\gamma$. We will consider only those automorphisms for which $T(\gamma)= \gamma$. Then $T(X)$, $T(Z)$ and $\gamma$ generate a group isomorphic to $\heis{N}$ provided $T(X)^N=T(Z)^N=1$. In particular if $c\in C(\heis{N})$ is invertible then the map $T(x)= cxc^{-1}$ is an automorphisms satisfying these conditions. Such automorphisms are called {\em inner}. Further call an inner automorphism unitary if $c^{-1}=c^*$ (see \eqref{eq:involution} for the definition of the $^*$ operation). We can prove the following. 
\begin{propn}
If $T(x)= cxc^{-1}$ is a unitary inner automorphism and $\phi$ is representation of $\heis{N}$ then there is a unitary operator $U_c$ such that $\phi(T(x))=U_c\phi(x)U_c^{-1}$. Conversely for any unitary operator $U$ on the representation space of $\heis{N}$ there is a unitary inner automorphism $T_U$ such that $U\phi(x)U^{-1}= \phi(T_U(x))$. Thus there is a one-to-one correspondence between the set $\cali{U}(\heis{N})$ of unitary inner automorphisms on $C(\heis{N})$ and quantum dynamics on the representative Hilbert space. 
\end{propn}
This result is neither difficult nor surprising given the fact that the $\heis{N}$ completely characterizes the kinematics of the system. It does however give us an alternative description and algebraic tools to study the dynamics. Thus we can study the effect of unitary operations on distribution functions\cite{Miquel} using these transformations. Note however that we allow {\em reducible} representations now. The set of automorphisms of the {\em group} $\heis{N}$ is a subgroup of $\cali{U}(\heis{N})$. 

In this work, we have concentrated on irreducible representations of $\heis{N}$ in which $\gamma$ acts maximally. By dropping the last assumption  we can get all finite dimensional representations. The order of $\phi(\gamma)$ in the representation $\phi$ is the dimension. We can then use the products of these representations (actually we need some extra structure) to study unitary gates. We aim to explore this in future. Let us note some interesting relations in the case $N=2^n$. If $u\in \heis{N}$ we will  denote by $\phi_k$ the representation in which $\gamma^{2^k}=1$. Let $\sigma_i,\; i=1,2,3$ denote the Pauli matrices and $I_r$ the identity matrix of order $r$. Then
\[
\begin{split}
& \phi_1(X)= \begin{pmatrix} 0& 1 \\1 & 0 \end{pmatrix} = \sigma_1\quad \phi_1(Z)= \begin{pmatrix} 1& 0 \\0 & -1 \end{pmatrix}=\sigma_3\\
& \phi_2(X)= C\sigma_1\tensor \sigma_1 \quad \phi_2(Z) = \sigma_3\tensor S \\
& \text{ where } C = \begin{pmatrix} 1& 0 & 0 & 0 \\ 0& 0 & 0 & 1 \\ 0 & 0 & 1 & 0\\ 1 & 0 & 0 & 0 \end{pmatrix} \text{ and } S =\begin{pmatrix} 1 & 0 \\ 0 & i \end{pmatrix} \\
\end{split}
\]
are a CNOT gate ($C$) and the phase gate ($S$) respectively \cite{NC}. We also note that in the general case $\phi_n(X)$ is the cyclic shift operator. It can be efficiently constructed, for example, using  full adder circuits  with $n+1$ ancillary qubits. Similarly, $\phi_n(Z)$ can be constructed using appropriate controlled phase gates as in the quantum Fourier transform. We also observe that iterating the simple relations $\phi_k(X^2)= \phi_{k-1}(X)\tensor I_2 $ and  $\phi_k(Z^2)= I_2 \tensor\phi_{k-1}(X) $ we obtain the interesting relations 
\beq\label{eq:RelnReps}
\phi_n(X^{2^k})= \phi_{n-k}(X)\tensor I_{2^k}\text{ and } \phi_n(Z^{2^k})= I_{2^k}\tensor \phi_{n-k}(Z)
\eeq
These relations can be used to devise more efficient implementations. 

We conclude this section with a discussion of potential application of these constructions to {\em quantum process tomography} \cite{Mohseni}. A quantum process is characterized by a completely positive map $T$ acting on the operators on the system Hilbert space. If we have a complete set of phase-point operators $\{A(x,z)\}$ then $T$ is determined by its action on these. 
Let us assume that the dimension is odd so that we have a  set of phase-point operators satisfying the full set of marginal conditions. Using Theorem \ref{thm:Radon1} we can prove the following. 
\begin{propn}
Let $T$ be quantum process (a CP map) given by 
\[ T(A(x,z)) = \sum_{x',z'} T(x',z':x,z) A(x',z')\]
Here $T(x',z':x,z)$ is the ``matrix'' of $T$ in the basis $\{A(x,z)\}$ of phase-point operators. 
\[ M^{-1} = \begin{pmatrix} a & b \\ c & d \end{pmatrix} \text{ and }M'^{-1} = \begin{pmatrix} a' & b' \\ c' & d' \end{pmatrix}\in SL(2,Z_N)\] 
Then writing ${\bf u} = (u_1, u_2)=(x,z)$ and  ${\bf u}' = (u_1', u_2')=(x',z')$
\[ \sum_{(M'^{-1}{\bf u}')_1, (M^{-1}{\bf u})_1} T(M'^{-1}{\bf u}':M^{-1}{\bf u}) = \inprab{z'+\frac{c'd'}{2}}{T\left(\pj{z+\frac{cd}{2}}\right)}\]
Here $\{\ket{z}\}$ and $\{\ket{z'}\}$ denote the {\em ordered} basis of eigenvectors of $\sigma_M(Z)$ and $\sigma_{M'}(Z)$ respectively. 
\end{propn}
We do not prove it here since it is similar to the proof given in Theorem \ref{thm:Radon1}. Note that we are averaging over two indices now. To use the theorem we apply $T$ to the projections $\pj{z}$ and measure the result in the basis $\{\ket{z'}\}$. These transition probabilities yield the right hand side in the above equations. In principle, these equations can be inverted using the inverse Radon transforms (see section \ref{sec:inverseRadon}) to yield the coefficients $ T(x',z':x,z)$ and thus determining $T$ (see \cite{Paz04} for a different perspective on phase-space tomography). Several optimizations are possible especially if we have some prior knowledge of the process. But we do not discuss these issues here as they merit a separate investigation. 
\section{Discussion}
In this work we have analyzed quasiprobability distribution functions corresponding to quantum states. Our viewpoint is that these are the real coefficients of bases (or generally frames) in the space of hermitian operators. The choice of these bases is dictated by certain conditions we impose. This leads to expressing these bases or collection of phase-point operators in terms of operators representing the Weyl-Heisenberg groups. In the language of frame theory \cite{Christensen03, Ferrie} these operators generate the Weyl-Heisenberg frames. We do not go into the intricacies of frame theory approach here. Our approach is more group-theoretic, emphasizing the role of Weyl-Heisenberg groups in quantum kinematics.  The other groups which play an important role are $SL(2,Z_N)$ which yield the marginals. Conversely, we can use distribution functions to study these groups. 
We have given explicit formulas for the Radon transforms and their inversions. These can be used to solve the problem state or operator reconstruction. Even when we do not have sufficient data on marginals to invert the transforms we can get partial information about the state by taking {\em generalized} inverses \cite{Fill,Velasquez}. We aim to address these and other issues on state and process estimation and reconstruction including the practical and computational aspects in future. 
\bibliographystyle{iopart-num}
\bibliography{FiniteWigner}

\providecommand{\newblock}{}
\begin{thebibliography}{10}
\expandafter\ifx\csname url\endcsname\relax
  \def\url#1{{\tt #1}}\fi
\expandafter\ifx\csname urlprefix\endcsname\relax\def\urlprefix{URL }\fi
\providecommand{\eprint}[2][]{\url{#2}}

\bibitem{Wigner32}
Wigner E~P 1932 {\em Phys. Rev.\/} {\bf 40} 749

\bibitem{Shiryayev}
Shiryayev A~N 1984 {\em Probability\/} (Springer-Verlag)

\bibitem{Moyal}
Moyal J~E 1949 {\em Proc. Camb. Phil. Soc.\/} {\bf 45} 99

\bibitem{Strat}
Stratonovich R~L 1957 {\em Sov. Phys. JETP\/} {\bf 4} 891

\bibitem{Rotman}
Rotman J~J 1994 {\em An Introduction to the Theory of Groups\/} 4th ed
  (Springer-Verlag)

\bibitem{Baker}
Baker G~A 1958 {\em Phys. Rev.\/} {\bf 109} 2198

\bibitem{Wootters87}
Wootters W~K 1987 {\em Ann. Phys. (NY)\/} {\bf 176} 1

\bibitem{Vourdas}
Vourdas A 2004 {\em Rep. Prog. Phys.\/} {\bf 67} 267

\bibitem{Gabor46}
Gabor D 1946 {\em Journ. I.E.E.\/} {\bf 93} 429

\bibitem{Christensen03}
Chistensen O 2003 {\em {An Introduction to Frames and Riesz Bases}\/}
  ({Birkh\"{a}ser})

\bibitem{Ferrie}
Ferrie C and Emerson J 2009 {\em New. J. Phys.\/} {\bf 11} 1

\bibitem{Durt}
Durt T, Englert B~G, Bengtsson I and \'Zyckowski K 2010 {\em Int. J. Quant.
  Inf.\/} {\bf 8} 535

\bibitem{Leonhardt95}
Leonhardt U 1995 {\em Phys. Rev. Lett.\/} {\bf 74} 4101

\bibitem{Leonhardt97}
Leonhardt U 1997 {\em {Measuring the quantum state of light}\/} (Cambridge
  University Press)

\bibitem{Perelomov}
Perelomov A 1984 {\em {Generalized coherent states and applications}\/}
  ({Springer-Verlag})

\bibitem{Miquel}
Miquel C, Paz J~P and Saraceno M 2002 {\em Phys. Rev. A\/} {\bf 65} 062309

\bibitem{Paz05}
Paz J~P, Roncaglia A~J and Saraceno M 2005 {\em Phys. Rev. A\/} {\bf 72} 012309

\bibitem{Gibbons}
Gibbons K~S, Hoffman M~J and Wootters W~K 2004 {\em Phys. Rev. A\/} {\bf 70}
  062101

\bibitem{Weyl}
Weyl H 1950 {\em {The theory of groups and quantum mechanics}\/} ({Dover})

\bibitem{Schwinger70}
Schwinger J 1970 {\em {Quantum Kinematics and Dynamics}\/} ({Benjamin})

\bibitem{Appleby}
Appleby D~M 2005 {\em J. Math. Phys.\/} {\bf 46} 052107

\bibitem{Terras}
Terras A 1999 {\em Fourier Analysis on Finite Groups\/} (CUP)

\bibitem{Folland89}
Folland G~B 199 {\em Harmonic Analysis in Phase Space\/} (Princeton University
  Press)

\bibitem{Gross}
Gross D 2006 {\em J. Math. Phys.\/} {\bf 47} 122107

\bibitem{Leonhardt96}
Leonhardt U 1996 {\em Phys. Rev. A\/} {\bf 53} 2998

\bibitem{Chaturvedi09}
Chaturvedi S, Mukunda N and Simon R 2009 {Wigner distributions for finite state
  systems without redundant phase point operators} arXiv:0909.1387v1 [quant-ph]

\bibitem{Diaconis}
Diaconis P and Graham R~L 1985 {\em Pacific. J. Math.\/} {\bf 118} 323

\bibitem{Fill}
Fill J~A 1989 {\em SIAM. J. Disc. Math.\/} {\bf 2} 262

\bibitem{Velasquez}
Velasquez E 1985 {\em Pacific. J. Math.\/} {\bf 177} 369

\bibitem{Arthurs65}
Arthurs E and J~L~Kelley J 1965 {\em Bell. Syst. Tech. J.\/} {\bf 44} 725

\bibitem{Braunstein91}
Braunstein S~L, Caves C~M and Milburn G~J 1991 {\em Phys. rev. A.\/} {\bf 43}
  1153

\bibitem{Hillery84}
Hillery M, O'Connell R, Scully M~O and Wigner E~P 1984 {\em Phys. Rep.\/} {\bf
  106} 121

\bibitem{Serre}
Serre J~P 1977 {\em {Linear Representations of Finite Groups}\/} (Springer)

\bibitem{Mordell69}
Mordell L~J 1969 {\em {Diophantine Equations}\/} ({Academic Press})

\bibitem{NC}
Nielsen M and Chuang I 2000 {\em {Quantum computation and quantum
  information}\/} ({CUP})

\bibitem{Mohseni}
Mohseni M, Rezakhani T and Lidar D~A 2008 {\em Phys. rev. A\/} {\bf 77} 032322

\bibitem{Paz04}
Paz J~P, Roncaglia A~J and Saraceno M 2004 {\em Phys. Rev. A\/} {\bf 69} 032312

\end{thebibliography}
\appendix
\section{Appendix}
\begin{proof}[Proof of Proposition \ref{thm:contWignerDist}]
We prove only \eqref{eq:Radon_marginal}. 
\commentout{
We first observe that the commutator 
\[ 
[\genX',\genZ']= [\cos{\theta} \genX +\sin{\theta}\genZ, -\sin{\theta} \genX+\cos{\theta}\genZ] 
= [\genX,\genZ]
\]
}
The invariance of scalar product and the standard measure on $\real^2$ under rotation implies that 
\[
\begin{split} 
& \int W_c(x,z:\rho){\mathrm d}x'= \int \diffD x'\int \ave{e^{ i(u\genX+v\genZ)}} e^{-i(ux+vz)} \diffD u \diffD v\\
& = \int \diffD x'\int \ave{e^{ i(u'\genX'+v\genZ')}} e^{-i(u'x'+v'z')}\diffD u' \diffD v' \\
& = \int \ave{e^{ i(u'\genX'+v\genZ')}} e^{-i(u'x'+v'z')}\delta(u')\diffD u' \diffD v' 
= \int \ave{e^{iv'\genZ'}} e^{-iv'z'} \diffD v'\\
\end{split}
\]
Let $\ket{z'}$ be the eigenvectors of $\genZ'$ with eigenvalues $z'$. Then 
\[ \int_a^b\diffD z'\ave{e^{iv\genZ'}}= \int_a^b \diffD z' e^{ivz'} \inpra{z'}{\rho} \]
and equation \eqref{eq:Radon_marginal} follows. 
\end{proof}
\begin{proof}[Proof of Proposition \ref{prop:Radon2}]
We will prove the second formula only. The proof is similar for the first formula. Using the induced automorphism given in \eqref{eq:induced_auto} we get
\[
\begin{split}
&\sum_{x'\in Z_n} W_0(M^{-1}x', M^{-1}z':\rho)=\\
& \frac{1}{N^2}\sum_{x'\in Z_N}\sum_{m,n} \omega^{(1-\delta_{am+cn,0})(1-\delta_{bm+dn,0})((a+b)m+(c+d)n)^2N/4}\omega^{(am+cn)(bm+dn)/2}\\
&\omega^{-(abm(m-1)/2+cdn(n-1)/2+cdmn)}\ave{\sigma_M(X)^{m}\sigma_M(Z)^{n}}\omega^{-(mx'+nz')}\\
&= \frac{1}{N}\sum_{x'\in Z_N}\sum_{m,n}\omega^{(1-\delta_{am+cn,0})(1-\delta_{bm+dn,0})((a+b)m+(c+d)n)^2N/4}\omega^{(am+cn)(bm+dn)/2}\\
&\omega^{-(abm(m-1)/2+cdn(n-1)/2+(m+n)(cd+\sgn(cd)/2))}\ave{\sigma_M(X)^{m}\sigma_M(Z)^{n}}\omega^{-nz'}\delta_{m0}\\
&= \frac{1}{N}\sum_{n} \omega^{((c+d)n)^2N/4}\omega^{(cn)(dn)/2}\omega^{-\sgn(cd)n/2-cdn(n-1)/2}\ave{\sigma_M(Z)^n}\omega^{-nz'}.\\
\end{split}
\]
Here we use that fact that $cn,dn\neq 0\bmod N$ for any odd $n$ since $N/u$ is even. We have to consider two cases separately suppose first that $cd$ is odd. Then 
\[
\begin{split}
&\sum_{x'\in Z_n} W_0(M^{-1}x', M^{-1}z':\rho)=\\
&=\frac{1}{N}\sum_j\rho_{jj}\Bigl(\sum_{r=0}^{N/2-1}\omega^{(cd/2+j-z')(2r)} +\\ &\sum_{r=0}^{N/4-1}(-1)^{d\bfloor{\frac{c(2r+1)}{N}}+c\bfloor{\frac{d(2r+1)}{N}}}\omega^{(cd/2+j-z')(2r+1)}+\text{comp. conj. }\Bigl)\\
&= (\inpra{z'-cd/2}{\rho}+\inpra{N/2+z'-cd/2}{\rho})/2\\
&+\frac{2}{N} \sum_{j}\rho_{jj}\sum_{\substack{n \text{ odd}\\ n <N/2}}(-1)^{d\bfloor{\frac{cn}{N}}+c\bfloor{\frac{dn}{N}}}\cos{\frac{2\pi(cd/2+j-z')}{N}}
\end{split}
\]
We can prove the second case ($cd$ odd) similarly. 
\end{proof}
\end{document}